\documentclass[preprint,12pt]{elsarticle}

\usepackage{amssymb}
\usepackage{graphicx}
\usepackage[usenames,dvipsnames]{color}
\usepackage{subfig,epstopdf}
\usepackage{mathtools}
\usepackage{dcolumn}
\usepackage{bm}
\usepackage{geometry}
\def \tlim {t}
\def \krondelta {\delta^{k}_{\phi(\omega),0}}
\def \delkpq {\delta_{{\bf k},{\bf p} {\bf q}}}
\def \delpkq {\delta_{{\bf p},{\bf k} {\bf q}}}
\def \delqpk {\delta_{{\bf q},{\bf p} {\bf k}}}

\def \delokpq {\delta_{\omega_\mathbf{k}, \omega_\mathbf{p} \omega_\mathbf{q}}}

\def \dpq {d {\bf p} d {\bf q}}
\def \dkpq {d {\bf k} {\bf p} {\bf q}}

\def \ck {c_{\bf k}} \def \cks {c^*_{\bf k}}
\def \cp {c_{\bf p}} \def \cps {c^*_{\bf p}}
\def \cq {c_{\bf q}} \def \cqs {c^*_{\bf q}}
\def \cm {c_{\bf m}}

\def \Ls {\tilde{L}} 

\def \G {\Gamma}

\journal{Physica D}

\begin{document}

\begin{frontmatter}



\title{Multiple scales kinetic theory for rotationally constrained slow inertial waves and anisotropic dynamics}


\author{Amrik Sen\footnote[2]{amriksen@gmail.com}}
\address{Department of Applied Mathematics, University of Colorado, Boulder, USA\\
and\\
Tata Institute of Fundamental Research, Hyderabad, India.
}

\begin{abstract}
Wave kinetic theory for rapidly rotating flows is developed in this paper using a rigorous application of multiple scales perturbation theory. The governing equations are an asymptotically reduced set of equations that are derived from the incompressible Navier-Stokes equations. These equations are applicable for rapidly rotating flow regimes and are best suited to describe anisotropic dynamics of rotating flows. The independent variables of these equations inherently reside in a helical wave basis that is the most suitable basis for inertial waves. A coupled system of equations for the two global invariants: \textit{energy} and \textit{helicity}, is derived by extending a simpler symmetrical system to the more general non-symmetrical helical case. This approach of deriving the kinetic equations for helicity follows naturally by exploiting the symmetries in the system and is different from the derivations presented in earlier work of \citet{Galtier03, Galtier14} that uses multiple correlation functions to account for the asymmetry due to helicity. Stationary solutions, including Kolmogorov solutions, for the flow invariants are obtained as a scaling law of the anisotropic wave numbers. The scaling law solutions compare affirmatively with results from recent experimental and simulation data. The theory developed in this paper pertains to the wave dynamics supported by an asymptotically reduced set of hydrodynamic equations and therefore encompasses a different dynamical regime compared to the weak turbulence theory presented in the work of \citet{Galtier03, Galtier14}.    
\end{abstract}


\end{frontmatter}



\section{INTRODUCTION}
\label{sec:intro}

The theory of weak wave dynamics (i.e., the stochastic theory of nonlinear wave interactions) has been extensively studied since the seminal works of \citet{Kadom64}, \citet{Galeev65}, \citet{Zakha67} and more recently reviewed by  \citet{Zakha97}, \citet{Balk00}, \citet{Choi04} and \citet{Naza11}. This theory remains one of the few areas where a rigorous mathematical framework exists with predictive capabilities for studying the energetics  and dynamics associated with fluid turbulence. In this regard, the theory of weak wave turbulence was further explored by  \citet{Galtier03, Galtier14} and  \citet{Cambon04} in the context of unstratified rotating  turbulent flows. The theory utilizes the incompressible Euler equation for inviscid dynamics in an infinite space, in dimensionless form:
\begin{equation}
(\partial_t + {\bf u}\cdot \nabla){\bf u} + \frac{1}{\mathcal{R}o}{\bf \hat{z}} \times {\bf u} = -\frac{1}{\mathcal{R}o}\nabla p, \quad \nabla \cdot {\bf u}=0.  \label{NSE}
\end{equation}
Here $\mathbf{u}=(u,v,w)$ is the three-dimensional velocity field in the Cartesian geometry $\mathbf{x}=(x,y,z)$, $p$ is the pressure field, and $\nabla=(\partial_x,\partial_y,\partial_z)$ is the gradient operator.
Equation (\ref{NSE}) is characterized by system rotation $2\Omega \mathbf{\widehat{z}}$ along with length, advective velocity, time and pressure scales respectively denoted by $L, U, \frac{L}{U}$ and $2\Omega L U$. The Rossby number $\mathcal{R}o=\frac{U}{2\Omega L}$, describing the relative importance of nonlinear advection to the Coriolis acceleration force, is the sole non-dimensional parameter. Of particular interest is the regime  $\mathcal{R}o\ll1$ for rotationally constrained flows along with the concomitant viewpoint that 
the dynamics can be partitioned into fast inertial waves  evolving on the  rotational timescale of ${\cal O}((2\Omega)^{-1})$, 
and eddies evolving on the advection timescale of ${\cal O}(\frac{L}{U})$. In nondimensional units these timescales are denoted by ${\cal O}(\mathcal{R}o)$ and ${\cal O}(1)$ respectively. In the Cartesian coordinate system
the linear inertial-wave dispersion relation, obtained from eq.~(\ref{NSE}) for  Fourier plane waves with wavevector  ${\mathbf k}=(k_x,k_y,k_z)$ and  of the form 
$\exp \{i ( {\mathbf k} \cdot {\bf x} - \omega_{\bf k} t )\}$,
is given by
\begin{equation}
\label{eqn:disp}
\omega_\mathbf{k} = \pm \frac{1}{\mathcal{R}o} \frac{k_z}{\sqrt{k_x^2 + k_y^2 + k_z^2}},
\end{equation}
Slow dynamics, to leading order, are geostrophically balanced, i.e., 
\begin{equation}
\label{geos}
{\bf \hat{z}} \times {\bf u} \approx - \nabla p.
\end{equation}
 It follows ${\bf u}_\perp \approx \nabla^\perp p$, with ${\bf u}_\perp = (u,v,0)$ and $\nabla^\perp=(-\partial_y,\partial_x,0)$, such that pressure is now identified as the geostrophic streamfunction. 
On noting $\nabla_\perp=(\partial_x,\partial_y,0)$ equation (\ref{geos}) implies that geostrophic motions are horizontally non-divergent with $\nabla_\perp\cdot{\bf u}_\perp\approx0$. Moreover, such motions
are columnar in nature due to the Taylor-Proudman constraint that enforces axial invariance, i.e.,  $\partial_z ({\bf u}, p) \approx 0$ \citep{jP16,gT23}.
Recent work \citep{kj06,Julien07} has more precisely established this invariance as true provided $k_z \gg  {\cal O} (\mathcal{R}o)$, thus providing an upper bound to the degree of
spatial anisotropy,  i.e.,  $\partial_z ({\bf u}, p) ={\cal O}(\mathcal{R}o)$. 

Underlying hypotheses for the theory of rotating  wave  turbulence are: (i) the separation of inertial and advective timescales, i.e. $\mathcal{R}o \ll 1$ and (ii) the non-interaction  between geostrophically balanced  and inertial waves dynamics\citep{EM98}. A necessary criterion for this to occur in eq.~(\ref{NSE})  is $(\vert {\bf u} \vert, p) =o(\mathcal{R}o^{-1})$ which ensures that the nonlinear terms remain small compared to linear terms. It follows from this bound that wave amplitudes can be significant. Within the $\mathcal{R}o\ll1$ regime, laboratory experiments  \citep{Thiele09, Morize06} and sufficiently spatially resolved simulations \citep{SW99, Cambon04, Bour07} have clearly demonstrated the tendency for the inertial wave spectra to evolve anisotropically towards a slow manifold associated with axially invariant  geostrophic dynamics ($k_z = 0$).  Using wave turbulence theory on equation \eqref{NSE}, \citet{Galtier03} 
predicts an anisotropic energy spectrum $E(k_\perp,k_z) \sim k_{\perp}^{-5/2}k_z^{-1/2}$ and a helicity spectrum $H(k_\perp,k_z) \sim k_{\perp}^{-3/2}k_z^{-1/2}$ contrary to predictions  of a $k_\perp^{-2}$ solution for $E_{k_\perp}$ observed in simulations \citep{MinPou10}, \citep{Teitel12}. A critical concern for this discrepancy is that the uniformity of the asymptotic approach of the weak-wave turbulence theory is lost as the slow manifold is approached. Notably as $k_z \to {\cal O}(\mathcal{R}o)$ it is found that inertial waves still exist within the slow manifold and are slow. Such \textit{slow} waves are not accounted for in \citet{Galtier03, Galtier14}.
 \begin{figure}[h!]
\begin{center}
\includegraphics[scale=0.8]{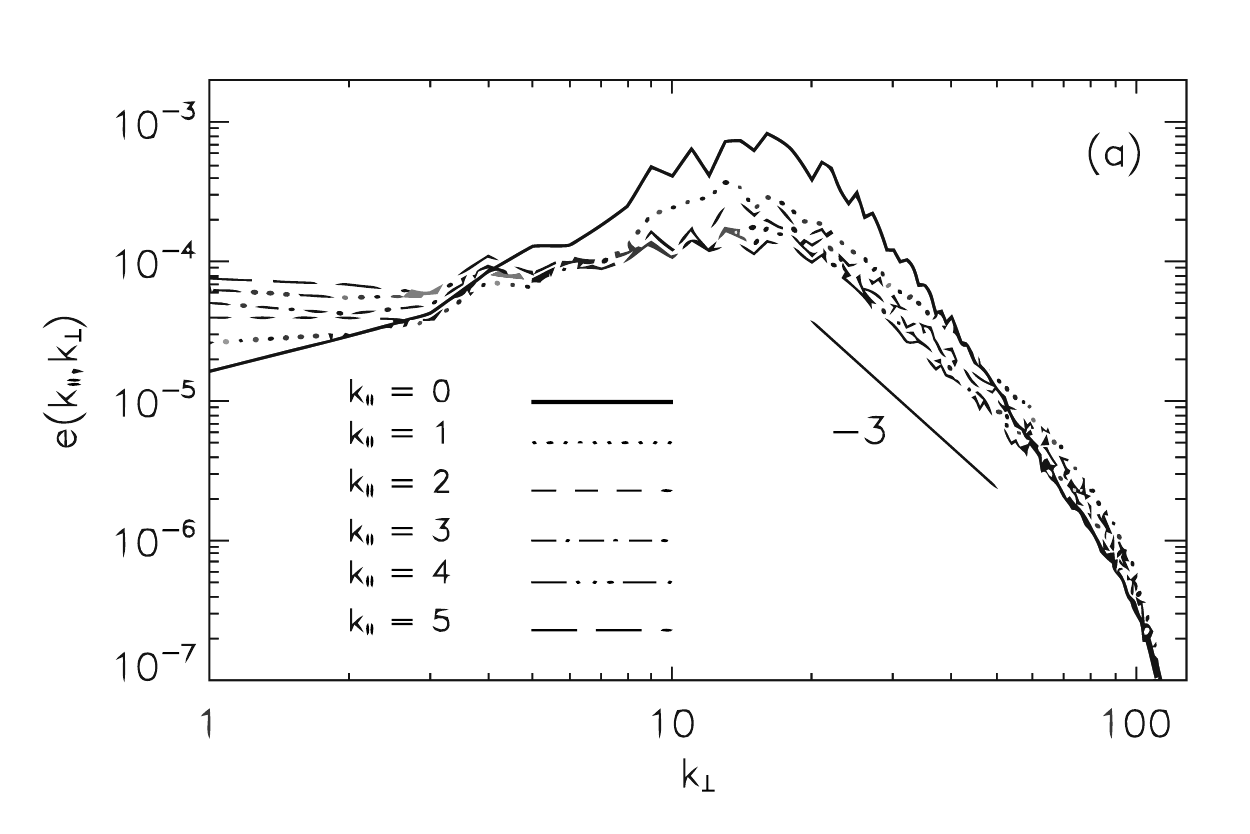}
\end{center}
\caption {Anisotropic energy spectra from rotating helical turbulence simulation\citep{Teitel12}. The $k^{-3}_\perp$ energy spectra shown here corresponds to the cylindrical symmetrical $k^{-2}_\perp$ energy spectra as is explained later in this paper.}
\label{Pablo_spec}
\end{figure}

In this paper, we apply a rigorous multi-scale perturbation method directly within the slow manifold to a recently derived set of asymptotically reduced equations for rotationally constrained flows \citep{JKW98,kj06,Julien07,NS11}, see section~\ref{R-RHD}. In the following, we adopt the nomenclature  \textit{reduced rotating hydrodynamic}  (R-RHD) for the reduced equations describing an unstratified, non-buoyant rotating fluid. We note that R-RHD capture a slow manifold that is more precisely identified as $k_z\sim {\cal O}(\mathcal{R}o)$ and contains not only geostrophic columnar eddies but also anisotropic  inertial waves characterized by scales: $k_z/k_\perp \ll \mathcal{R}o \ll 1$. The amplitude of these \textit{slow} waves evolve on slower advective timescale.  

For convenience, to aid our investigation, we centrally locate some nomenclature involving position vector ${\mathbf x}$ and wavenumber vector ${\mathbf k}$, namely
\begin{eqnarray}
{\mathbf x} = {\mathbf x}_\perp + Z{\mathbf{\widehat z}}, &\qquad & {\mathbf k} = {\mathbf k}_\perp + k_Z{\mathbf{\widehat z}}\\
\quad {\mathbf x}_\perp = (x,y,0), &\qquad &  {\mathbf k}_\perp = (k_x,k_y,0) \\
 {\mathbf x}^\perp = (-y,x,0), &\qquad & {\mathbf k}^\perp = (-k_y,k_x,0)\\
\nabla_\perp = (\partial_x, \partial_y, 0), &\qquad & \nabla^\perp = (-\partial_y, \partial_x, 0)
\end{eqnarray}
with $\vert{\mathbf k}_\perp \vert = \vert {\mathbf k}^\perp\vert = k_\perp=\sqrt{k_x^2 + k_y^2}$. Once we introduce the helical basis in section \ref{HELBASIS}, we note that the position vector and the wave vector will be defined in the appropriate right hand coordinate frame of the helical basis as follows: $({\mathbf x}^\perp, \widehat{{\bf z}}, {\bf x}^\prime)$ and $({\mathbf k}^\perp, \widehat{{\bf z}}, {\bf k}^\prime)$ respectively. Here, prime refers to a vector antiparallel to the parent vector, i.e. ${\bf k^\prime} = -{\bf k}$. The superscript prime will be simply dropped when writing the wave vector. 

\section{Reduced-Rotating Hydro-Dynamic Equations, R-RHD}
\label{R-RHD}

The detailed derivation of R-RHD is provided in \citet{kj06,Julien07}. To summarize, the asymptotic framework for equation (\ref{NSE}) is established by assuming the small expansion parameter $\mathcal{R}o$ and a multiple-scale expansion in 
the axial direction  $\partial_z = \partial_\mathrm{z}  + \mathcal{R}o \partial_Z$  with the isotropic scale $\mathrm{z}=z \sim (x,y)$ and  
the anisotropic columnar length scale $Z =\mathcal{R}o z$.  
Fluid variables ${\bf v} = ({\bf u}, p)^T$, where $T$ denotes tranpose, are now written as an asymptotic series in terms of the small parameter, $\mathcal{R}o$:
\begin{equation}{\bf v} = {\bf v_0} + \mathcal{R}o {\bf v_1} + \mathcal{R}o^2 {\bf v_2} + \mathcal{O}(\mathcal{R}o^3).\end{equation}
To leading order in equation (\ref{NSE}), we observe a point wise geostrophic balance: ${\bf\widehat z}\times {\bf u}_0 = -\nabla p_0$. It follows that fluid motions are horizontally non-divergent, i.e. $\nabla_\perp \cdot {\bf u}_{0\perp} = 0$, with ${\bf u}_{0\perp} = \nabla^\perp \psi$ where  $p_0=\psi$ is the geostrophic stream function as in the classical theory of quasigeostrophy.
The Taylor-Proudman constraint \citep{Gspan68} associated with the geostrophic balance further requires vertical variations to be negligible on ${\cal O}(1)$ vertical scales, i.e., $\partial_\mathrm{z} {\bf v}_0\equiv 0$.  Importantly, in compliance with the Taylor-Proudman constraint, the multiple scales approach permits variations of ${\cal O}(\mathcal{R}o^{-1})$ on the $Z$-scale \citep{kj06,Julien07}.
Hereafter, in the following we set $k_z=\mathcal{R}o k_Z$.

At the next order in equation (\ref{NSE}) the requisite solvability conditions, ensuring non-secular behavior of ${\bf v_1}$, lead to the R-RHD equations describing the evolution of unstratified slow geostrophically balanced  motions:
\begin{equation}
\partial_t \zeta + {\bf u}_{\perp} \cdot \nabla_\perp\zeta = \partial_Z W \label{R-RHD1a}
\end{equation}
\begin{equation}
\partial_t W + {\bf u}_{\perp} \cdot \nabla_\perp W = -\partial_Z  \psi \label{R-RHD1b}
\end{equation}
Here ${\bf u}_\perp = \nabla^{\perp}\psi$, and $\zeta := \nabla_{\perp}^2 \psi$ and $W$ are the $\mathbf{\widehat z}$ components of the vorticity  and velocity fields.
Akin to classical quasigeostrophic theory, nonlinear vertical advection, $W\partial_Z$, is an asymptotically subdominant process and does not appear.
However, unlike quasigeostrophic theory the velocity field is isotropic in magnitude with $\vert {\bf u}_\perp\vert \sim \vert W \vert$, 
hence the appearance of a prognostic equation for $W$. Physically, the R-RHD \eqref{R-RHD1a} and \eqref{R-RHD1b} state that unbalanced vertical pressure gradients drive vertical motions that are materially advected in the horizontal, in turn, vortical stretching due to vertical gradients in $W$ produce
vortical motions. The vertical velocity, $W$, also generates an ageostrophic velocity fleld, ${\bf u}^{ag}_\perp$, such that incompressibility, ${\nabla_\perp\cdot{\bf u}^{ag}_{1\perp}} +\partial_Z W=0$, holds to ${\cal O}(\mathcal{R}o)$.
The R-RHD remain valid provided $(\vert {\bf u}_\perp\vert, \vert W \vert) = o(\mathcal{R}o^{-1})$.
Consistent with the Euler equation (\ref{NSE}), the R-RHD  also conserve, in time, the volume-averaged kinetic energy ${\cal E}_V$ and 
helicity ${\cal H}_V$:
\begin{equation}
{\cal E}_V \equiv \int {\bf u\cdot u} dV= \frac{1}{2}\int \left (  \vert\nabla_\perp \psi\vert^2 + W^2 \right )dV,
\end{equation}
\begin{equation}
{\cal H}_V \equiv \int {\bf u\cdot} \nabla\times{\bf u} dV = 2 \int \left (  W \zeta  \right )dV.\hspace{1.5em} 
\end{equation}
In what follows, an investigation of wave dynamics will be applied to the R-RHD.

\subsection{Geostrophic inertial waves and eddies}
\label{WAVEFIELD}
We observe that upon linearization the R-RHD
support \textit{slow geostrophically balanced} inertial waves of the form,
\begin{align} 
\label{eqn:inertial}
\mathbf{\Psi}^{s_k}_{\bf k}e^{i \mathbf{\Phi}(\mathbf{k}, s_k \omega_{\bf k} )},\quad {\omega}_{\bf k} = \frac{k_Z}{ k_\perp}
\end{align} 
with planar phase function $\mathbf{\Phi}(\mathbf{k}, s_k \omega_{\bf k} ) =  (\mathbf{ k_\perp\cdot x_\perp} + k_Z Z- {s_k} \omega_{\bf k} t) = (\mathbf{ k^\perp\cdot x^\perp} + k_Z Z- {s_k} \omega_{\bf k} t)$ (cf. equation \eqref{eqn:disp}). When $k_Z=0$ this expression represents vertically invariant  modes with $\omega_{k} = 0$ (i.e. the 2D modes of turbulent eddies).
The circularly polarized wave amplitude vector is  
\begin{equation}
\label{eqn:eigv}
\mathbf{\Psi}^{s_k}_{\bf k}\equiv
\left(
\begin{array}{c}
\psi^{s_k}_{\bf k}     \\   W^{s_k}_{\bf k} 
\end{array}
\right)
= 
\left(
\begin{array}{c}
s_k /k_\perp   \\   1
\end{array}
\right)
c^{s_k}_{\bf k},
\end{equation}
where $s_k=\pm$ denotes the handedness, `$+$' for right-handed circularly polarized waves (with positive helicity) and `$-$' for left-handed circularly polarized waves (with negative helicity). Here, $c^{s_k}_{\bf k}$ is a complex amplitude function.

\subsection{Helical basis for circularly polarized inertial waves}
\label{HELBASIS}

\begin{figure}[h!]
\begin{center}
\includegraphics[scale=0.75]{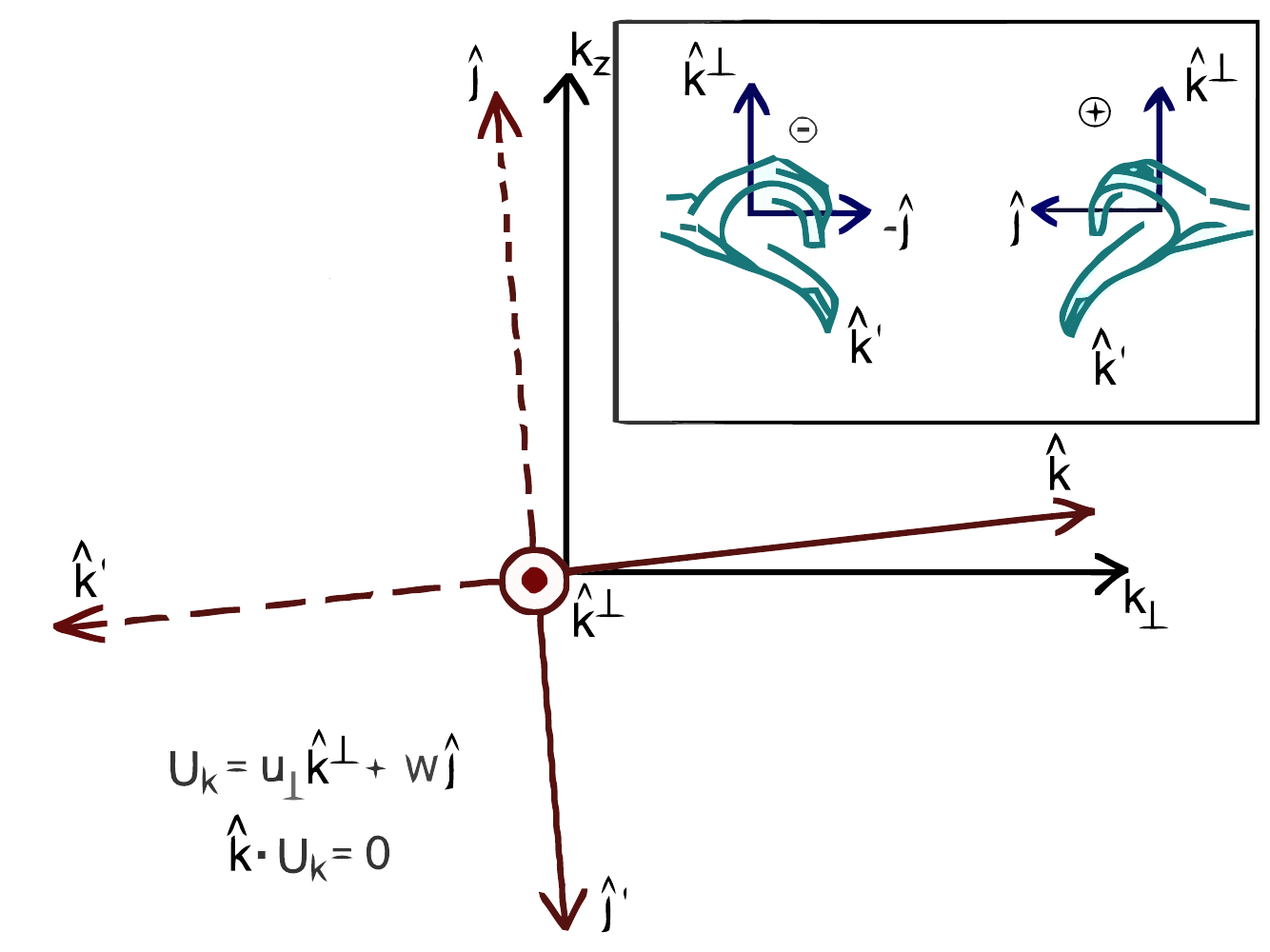}
\end{center}
\caption {\small{Helical wave basis: 
$({\mathbf{\widehat{k}}}^{\perp}, \boldsymbol{\widehat{\jmath}},{\mathbf{\widehat{k}}^\prime})$
forms a right-handed coordinate system with $\langle \widehat{\mathbf{k}}^\prime,\widehat{\mathbf{k}}^{\perp}\rangle= \langle \widehat{\mathbf{k}}^\prime,\boldsymbol{\widehat{\jmath}}\rangle=0$ where 
$\boldsymbol{\widehat{\jmath}}=\frac{{\mathbf{{\mathbf{{k}}^\prime} \times {k}}}^{\perp}}{k^2_\perp}$. The wave propagation direction is given by the wave vector, ${\bf \hat{k}}^\prime$ (which we simply call ${\bf k}$ in the body of the manuscript). Within the \textit{slow manifold} where $k_z =  {\mathcal{R}}o k_Z$, $({\mathbf{\widehat{k}}}^{\perp}, \boldsymbol{\widehat{\jmath}},{\mathbf{\widehat{k}}}^\prime)\rightarrow({\mathbf{\widehat{k}}}^{\perp}, \mathbf{\widehat{z}},{\mathbf{\widehat{k}}^\prime_\perp})$.}}
\label{hel_basis}
\end{figure}

We note that in wavenumber space the unit vectors $(\mathbf{\widehat k}^\perp,\mathbf{\widehat z},\mathbf{\widehat k}^\prime_\perp)$, with $\mathbf{\widehat k}^\prime_\perp = \mathbf{k^\prime}_\perp/k_\perp$ and  $\mathbf{\widehat k}^\perp = \mathbf{k}^\perp/k_\perp$,
form a right-handed orthogonal basis. Within the slow manifold, we have $\mathbf{\widehat k}^\prime_\perp \leftrightarrow \mathbf{\widehat k}^\prime$ as the direction of wave vector and henceforth we will simply call this vector ${\bf k}$ (see FIG. \ref{hel_basis}). The leading order velocity field associated with an inertial wave is given by 
\begin{eqnarray}
\mathbf{u} &=& \mathbf{u}_\perp + W \mathbf{\widehat z} \nonumber \\&=& \nabla^\perp \psi + W \mathbf{\widehat z} \\
&\xrightarrow{F.T.}& \left ( i k_\perp \psi^{s_k}_\mathbf{k} \mathbf{\widehat k}^\perp + W^{s_k}_\mathbf{k} \mathbf{\widehat z} \right)  e^{i \mathbf{\Phi}(\mathbf{k}, s_k\omega_{\bf k} )} + c.c. \nonumber \\
&=& \left ( U^{s_k}_\mathbf{k} \mathbf{\widehat k}^\perp + W^{s_k}_\mathbf{k} \mathbf{\widehat z} \right)  e^{i \mathbf{\Phi}(\mathbf{k}, s_k\omega_{\bf k} )} + c.c. \nonumber \\
&=& c^{s_k}_{\bf k} {\bf h}^{s_k}_{\bf k} e^{i \mathbf{\Phi}(\mathbf{k}, s_k\omega_{\bf k} )} + c.c. \nonumber
\end{eqnarray}
where $U^{s_k}_\mathbf{k}  := i k_\perp\psi^{s_k}_\mathbf{k}$. From \eqref{eqn:eigv}, ${\bf U}^{s_k}_{\bf k}:=(U^{s_k}_\mathbf{k}, W^{s_k}_\mathbf{k})^T= (i s_k, 1)^T={\bf h}^{s_k}_{\bf k}$ where 
\begin{equation}
\label{hel_basis0}
{\bf h}^{s_k}_{\bf k} := is_k \widehat{\bf k}^\perp + \mathbf{\widehat z}
\end{equation}
represents the complex helical wave basis (see ref. \citep{Lesieur72} for details) within the slow manifold incorporating the leading order incompressibility criteria $\nabla_\perp\cdot\mathbf{u}=0$, i.e. $\mathbf{ k}_\perp\cdot\mathbf{h}^{s_k}_{\bf k}=0$. Notably, as with its counterpart that exists outside the slow manifold (ref. Fig.~\ref{hel_basis}), this wave basis exhibits the following property that enables switching across different handedness by a conjugation operation,
\begin{equation}
\label{hel_bas_prop1}
{\bf h}^{-s_k}_{\bf k} = {\bf h}^{s_k*}_{\bf k}.
\end{equation}
These findings illustrate that the R-RHD are naturally set up in the helical wave coordinate basis.  

\section{Wave Amplitude Equations}
\label{WAMP}


In the framework of this paper we consider small amplitude dynamics. However, by comparison, we have established that the theory permits significant wave amplitudes of $o(\mathcal{R}o^{-1})$ outside the slow manifold. For connectivity and
consistency with observed energy spectra, one can therefore envision the scenario whereby the amplitudes of resonantly cascading inertial waves have been sufficiently attenuated once they reach the slow manifold. We therefore proceed with a multiple scales asymptotic approach in time, where $ \partial_t \to \partial_t  + \epsilon \partial_\tau$ and
\begin{align} 
{\bf \Psi}^\epsilon ({\bf x}_\perp,Z,t,\tau) &=  \epsilon {\bf \Psi}_1 ({\bf x}_\perp,Z,t,\tau) + \epsilon^2 {\bf \Psi}_2 ({\bf x}_\perp,Z,t,\tau) + \mathcal{O}(\epsilon^3); 
\label{amp_expansion}
\end{align}
where ${\bf \Psi_j}, ~\forall j = 1,2,3 ...$ is $\mathcal{O}(1)$ to ensure consistency of order in the expansion above. 

The order parameter $0< \epsilon\ll 1$  is a measure of the wavefield amplitude and  $\tau = \epsilon t$ denotes the slow advective timescale for \textit{weak} amplitude modulations in comparison to the inertial wave propagation time $t$. We note that $\epsilon\partial_\tau \sim \epsilon\mathbf{u}_1 \cdot \nabla_\perp$ indicating that $\tau$ defines the advective timescale. This ensures a necessary separation of temporal scales $\tau$ and $t$ as explained in \citep{EM98}, thereby allowing for a multi-scale treatment of the system with distinct dynamics at every order. The choice of an order parameter, $\epsilon$ proportional to ${k_Z}/k_\perp$ has been mentioned by \citet{NS11} and serves as a guiding principle for the theory developed here. This way a systematic multi-scale treatment that captures the anisotropy in the rotating system is developed. 

The leading order solution can be interpreted as a  complex field of waves and eddies undergoing resonant interactions on slow inertial timescale. At leading order, utilizing eq.~\eqref{hel_basis0}, we have that planar inertial waves satisfy
\begin{eqnarray}
{\cal L_H} \mathbf{U}^{s_k}_{1\bf k}
 = {\bf 0},
\qquad
{\cal L_H} \equiv \left [   - i  \omega^{s_k}_{\bf k} {\bf I}_2 - \frac{k_Z}{k_\perp} {\bf J}_2
\right ].
\label{R-RHDF2}
\end{eqnarray}
					      Here ${\bf J}_2 =\scriptscriptstyle{
\left(
\begin{array}{rr}
0  & 1     \\
 -1 &  0     
\end{array}
\right)}$ is the Hamiltonian matrix and ${\bf I}_2$ is the identity matrix. 
The solution to this system is the helical base vector ${\bf U}^{s_k}_{1\bf k}\equiv{\bf h}^{s_k}_{\bf k}$. It is important to note that ${\cal L_H}$ being non-Hermitian, an extended eigen basis, including both the left and right eigenvectors, is required. This extended basis is the complex helical basis $h^{s_k}_{\bf k}$ introduced earlier in eq.~(\ref{hel_basis0}).

Also of importance to the application of a solvability condition at higher asymptotic orders is the solution 
${\bf h}^{-s_k}_{\bf k}$ satisfying the adjoint problem 
$({\cal L_H})^{*T}{\bf h}^{-s_k}_{\bf k}\equiv {\bf h}^{-s_k T}_{{\bf k} }{\cal L_H} ={\bf 0}$ 
and orthogonality condition $\langle \frac{1}{2}{\bf h}^{-s_k}_{\bf k},{\bf h}^{s_k}_{\bf k} \rangle= 1$.
Given system (\ref{R-RHDF2}), a complex wave field  can now be expressed as a superposition of inertial waves and eddies (2D modes corresponding to $k_Z = 0$) represented in terms of the helical basis: 
\begin{equation}
\label{wave_hel_bas2}
{\bf U}_1=\sum_{s_k}\int \bigl \{c^{s_k}_{\bf k}(\tau){\bf h}^{s_k}_{\bf k}e^{i \mathbf{\Phi}(\mathbf{k}, s_k \omega_{\bf k} )} + c.c. \bigr \}d\mathbf{k}. 
\end{equation}
Here, $c^{s_k}_{\bf k}(\tau)$ is a complex amplitude function varying in the slow advective timescale, $\tau$. 

At  ${\cal O}({\epsilon^2})$, the next asymptotic order, we have for each helical mode $\mathbf{h}^{s_k}_\mathbf{k}$					              
\begin{align}					              
& \overbrace{{\cal L_H}{\bf U}^{s_k}_{2\bf k}}^{\text{1st term}}   
=- \overbrace{\partial_\tau  c_{\bf k}^{s_k} \mathbf{h}^{s_k}_\mathbf{k}}^{\text{2nd term}}
					        -\overbrace{\sum_{s_p,s_q}\int \frac{{\bf p}^{\perp} \cdot {\bf q}_{\perp}}{p_{\perp}} 
					      \left ( \begin{array}{r}
					               \frac{q_{\perp}}{k_{\perp}}U^{s_p}_{\bf p}U^{s_q}_{\bf q}\\
					               U^{s_p}_{\bf p}W^{s_q}_{\bf q}
					              \end{array} \right )
					               e^{i  \left( 
					               \mathbf{\Phi}(\mathbf{p}, s_p \omega_{\bf p} t)+
					               \mathbf{\Phi}(\mathbf{q},s_q \omega_{\bf q} t)
					               -\mathbf{\Phi}(\mathbf{k}, s_k \omega_{\bf k} t)
					               \right )
					               } 
					              \dpq}^{\text{3rd term}}   
					               \label{R-RHD_ham0}
\end{align}
where the subscript $1$ has been dropped from the right hand side of the above equations.  
Application of the solvability condition, $\frac{1}{2}({\bf h}^{-s_k}_{\bf k} \cdot {\cal L_H}{\bf U}^{s_k}_{2\bf k})= 0$, for bounded  growth in
${\bf U}^{s_k}_{2\bf k}$ then gives the wave amplitude equation
\begin{align}
\label{wave_amplitude1}
i\partial_\tau c^{s_k}_{\bf k}  = \frac{1}{2}\sum_{s_p,s_q}\int V^{s_ks_ps_q}_{kpq}c^{s_p}_{\bf p}c^{s_q}_{\bf q} 
\delkpq\krondelta \dpq, 
\end{align}
where $V^{s_ks_ps_q}_{kpq} :=  \frac{{\bf p}^{\perp} \cdot {\bf q}_{\perp}}{p_{\perp}} \left (\frac{q_\perp}{k_\perp}s_k s_p s_q + s_p  \right)$ is the interaction coefficient (\textbf{ref. Appendix \ref{freq_delta_amp_eq}}). Here $\krondelta$ is the Kronecker delta function and $\phi(\omega) = (s_k \omega_k - s_p\omega_p -s_q\omega_q)$. The conservation laws of the three wave dispersive system is given by the resonance condition that is succinctly captured by the two delta functions on the right hand side of the wave amplitude equation (\ref{wave_amplitude1}). The natural appearance of the two delta functions within the framework of the perturbation method presented here automatically guarantees that the conservation laws are not violated and also ensure that the triadic resonance condition plays the pivotal role in the wave dynamics. Note that the inner-product mentioned above involves a time integration of an exponential term over the large inertial time span which manifests as a frequency resonance condition thereby averting \textit{secular} terms in the perturbation analysis. This step is elaborated in \textbf{Appendix \ref{freq_delta_amp_eq}}. Resonating inertial wave modes form a complete set for all modes within the slow manifold. 

Imposition of eq.~(\ref{wave_amplitude1}) in eq.~(\ref{R-RHD_ham0}) implies that perturbed wavefield ${\bf U}^{s_k}_{2\bf k}$ is bounded and contains nonresonant waves whose wave amplitudes and energy are  by design small compared to resonant inertial waves, we will return to this point again in a latter section when we discuss coupling between the wave and 2D manifolds. Also for $s_k=+$, we note that the sum is actually carried over the combinations given by the set $(s_p,s_q) = \{(+,+),(+,-),(-,+)\}$ 
as resonance cannot be achieved for the $(-,-)$ case. The converse holds for $s_k=-$.

\subsection{Velocity spectral tensor}
\label{tensor}
Having deduced the wave amplitude equation \eqref{wave_amplitude1} we now proceed with the primary objective of finding the functional forms for stationary energy and helicity spectra, often referred to as Kolmogorov solutions \citep{Balk90}. The relation between energy and helicity in terms of the canonical complex variable, $c^{s_k}_{\bf k}$ can be understood by carefully analyzing the velocity spectral tensor in each spectral mode $\mathbf{k}$ . The velocity spectral tensor is defined as follows \citep{Les08}:
\begin{align}
\label{U_tensor1}
& \frac{1}{2}{\mathcal{U}_{\bf k}} := \frac{1}{2}\left ( {\bf U}^{}_{\bf k} \otimes {\bf U}^{*}_{\bf k'} \right)\delta_{\bf k',\bf k} \nonumber \\ & = \frac{1}{2}\left (\begin{array}{cc}
\sum\limits_{s_k = \pm}c^{s_k}_{\bf k}c^{s_k*}_{\bf k'}  & \sum\limits_{s_k = \pm}is_k\frac{k_{\perp}}{k_{\perp}}c^{s_k}_{\bf k}c^{s_k*}_{\bf k'}  \\
\sum\limits_{s_k = \pm}-is_k\frac{k_{\perp}}{k_{\perp}}c^{s_k}_{\bf k}c^{s_k*}_{\bf k'} &  \sum\limits_{s_k = \pm}c^{s_k}_{\bf k}c^{s_k*}_{\bf k'}    
\end{array}
\right)\delta_{\bf k',\bf k} \nonumber \\
& = \frac{1}{2}\left(\begin{array}{cc}
c^{+}_{\bf k}c^{+*}_{\bf k'}  + c^{-}_{\bf k}c^{-*}_{\bf k'}& \frac{ik_{\perp}}{k_{\perp}}(c^{+}_{\bf k}c^{+*}_{\bf k'}  - c^{-}_{\bf k}c^{-*}_{\bf k'})\\
\frac{-ik_{\perp}}{k_{\perp}}(c^{+}_{\bf k}c^{+*}_{\bf k'} - c^{-}_{\bf k}c^{-*}_{\bf k'})&  c^{+}_{\bf k}c^{+*}_{\bf k'} + c^{-}_{\bf k}c^{-*}_{\bf k'}    
\end{array}
\right)\delta_{\bf k',\bf k} \nonumber \\
& \xrightarrow{\int dk'} \frac{1}{2}\left(\begin{array}{cc}
c^{+}_{\bf k}c^{+*}_{\bf k}  + c^{-}_{\bf k}c^{-*}_{\bf k}& \frac{ik_{\perp}}{k_{\perp}}(c^{+}_{\bf k}c^{+*}_{\bf k}  - c^{-}_{\bf k}c^{-*}_{\bf k})\\
\frac{-ik_{\perp}}{k_{\perp}}(c^{+}_{\bf k}c^{+*}_{\bf k} - c^{-}_{\bf k}c^{-*}_{\bf k})&  c^{+}_{\bf k}c^{+*}_{\bf k} + c^{-}_{\bf k}c^{-*}_{\bf k}    
\end{array}
\right) \nonumber \\
& = \frac{1}{2}\left( \begin{array}{cc}
e^+_{\bf k} + e^-_{\bf k} & i\frac{h^+_{\bf k} + h^-_{\bf k}}{k_\perp}\\
-i\frac{h^+_{\bf k} + h^-_{\bf k}}{k_\perp} & e^+_{\bf k} + e^-_{\bf k}
\end{array}
\right) = \frac{1}{2}\left( \begin{array}{cc}
e_{\bf k}  & i\frac{h_{\bf k}}{k_\perp}\\
-i\frac{h_{\bf k}}{k_\perp} & e_{\bf k} 
\end{array}
\right). 
\end{align}
The Dirac delta function must be interpreted as a selector function, i.e. $\int f(x') \delta (x'-x) dx' = f(x)$. Consequently we have ensemble average of delta correlated terms given by $\langle c^+_{\bf k}c^{+*}_{\bf k'}\rangle := \int e^+_{\bf k'}\delta({\bf k'} - {\bf k}) d{\bf k'} = e^+_{\bf k} \equiv \int c^+_{\bf k}c^{+*}_{\bf k'}\delta({\bf k'} - {\bf k})d{\bf k'} = c^+_{\bf k}c^{+*}_{\bf k}$ (and similarly for $s_k = -$) which is used to write the second last equality in \eqref{U_tensor1} above.\footnote{Often the ensemble average is defined as $\langle c^+_{\bf k}c^{+*}_{\bf k'}\rangle = e^+_{\bf k'}\delta({\bf k'} - {\bf k})$ to reflect the homogeneity assumption\citep{Newell11}; however, here we are interested in the net average effect of the delta correlated terms in wavenumber space and hence it is perfectly alright to integrate this term.} This allows us to define the statistical quantities in the tensor relation (\ref{U_tensor1}) as follows: $e_{\bf k} := \sum_{s_k} \ck^{s_k} c^{s_k*}_{\bf k} = \sum_{s_k}e^{s_k}_{\bf k}$ and $ h_{\bf k} := \sum_{s_k} s_k k_\perp e^{s_k}_{\bf k}$. Note that the ensemble averaging is basically an integration operation over all possible wave numbers and only the term associated with the non-trivial value of the delta function survives. So arguments and results that follow in the rest of the paper must be interpreted in the statistical sense. From the relations in \eqref{U_tensor1}, it is clear that the following is true,
\begin{align}
e^+_{\bf k} + e^{-}_{\bf k} &= e_{\bf k}, \\
e^+_{\bf k} - e^{-}_{\bf k} &= \frac{h_{\bf k}}{k_\perp}.
\end{align}
By solving the above equations, we get,
\begin{align}
e^+_{\bf k} = \frac{1}{2} \biggl( e_{\bf k} + \frac{h_{\bf k}}{k_\perp}\biggr), \nonumber \\
e^-_{\bf k} = \frac{1}{2} \biggl( e_{\bf k} - \frac{h_{\bf k}}{k_\perp}\biggr). \label{eh_coupled_1}
\end{align}
These expressions are particularly useful in the derivation of the stationary energy and helicity spectra.

\section{ZERO HELICITY DYNAMICS}
\label{ZeroHel}
In this section, we  analyze the special case of a flow with zero helicity, i.e., $h_{\bf k} \equiv 0$ for all $\mathbf{k}$ such that  $e^+_{\bf k} = e^-_{\bf k}=\frac{1}{2}e_{\bf k}$. 
From \eqref{U_tensor1} this imposes the constraint on the complex amplitude functions $c^{-s_k}_{\bf k} = c^{s_k*}_{\bf k}$ where for sake of brevity we will write $c^{+*}_{\bf k}\equiv c^*_{\bf k} = c^-_{\bf k}$. This entails a reflection symmetry of the wave field and a \textit{reduction} in the Hamiltonian description of the system because a unique \textit{handedness}, associated with one of the $s$ variables, now describes the full system as the system described by $s=+$ and $s=-$ are mirror replicas of one another in the statistical sense, this point is explained in more detail in \citet{Sen_th}. \citet{Newell11} have exploited this reduction and have studied resonance wave dynamics for the nonlinear Schr\"{o}dinger's system.  
   
Now, the inertial waves occur in helicity couplets involving $\mathbf{h}^{+}_\mathbf{k}$ and $\mathbf{h}^{-}_\mathbf{k}=\mathbf{h}^{+*}_\mathbf{k}$ with wavefield given by 
\begin{equation}
\label{wave_hel_bas3}
{\bf U}_1= \int \bigl \{\left ( c^+_{\bf k}(\tau)\mathbf{h}^{+}_\mathbf{k}e^{-i\omega_{\bf k}t}  + 
c.c.
 \right )
e^{i (\mathbf{k}_\perp\cdot\mathbf{x}_\perp + k_Z Z)} + c.c. \bigr \}d\mathbf{k}. 
\end{equation}

\subsection{The Hamiltonian}
The Hamiltonian for the R-RHD can be expressed as a power series of the complex amplitudes, $\ck$, as follows:
\begin{equation}
H = H^{(3)} + H^{(4)}+\cdots \label{ham_ser}.
\end{equation}
The $H^{(4)}$ component denotes four-wave resonant interactions which are negligibly small and therefore not considered.
Resonant self-interactions, captured by $H^{(2)} := \int \omega_k \ck \cks d{\bf k}$, evolve on the inertial timescale $t$ and is already accounted for within  the operator ${\cal L_H}$. 
The wave amplitude equation \eqref{wave_amplitude1} may be re-expressed in terms of the Hamiltonian according to 
\begin{equation}
\label{Ham_eqn_0}
i\partial_\tau \ck = \frac{\delta H}{\delta \cks},
\end{equation}
where the leading order Hamiltonian $H\approx H^{(3)}$ is constructed by multiplying equation~(\ref{wave_amplitude1}) by $\ck^{*}$ and subtracting it from it's conjugate counterpart and integrating over the wave number space. Concomitant with the construction of the Hamiltonian emerges an evolution equation for the energy spectral density $e_{\bf k}$. This indicates that $H^{(3)}$ is a real-valued symmetric cubic functional in $\ck$. Upon symmetrization of a trivariate function $\G({\bf k},{\bf p},{\bf q})$,
\begin{align}
Sym\{ \G({{\bf k}, {\bf p}, {\bf q}})\} &= \frac{1}{6} \biggl\{ \G({{\bf k}, {\bf p}, {\bf q}}) +  \G({{\bf k}, {\bf q}, {\bf p}})  +  \G({{\bf p}, {\bf k}, {\bf q}}) + \G({{\bf p}, {\bf q}, {\bf k}})  +  \G({{\bf q}, {\bf k}, {\bf p}})  +  \G({{\bf q}, {\bf p}, {\bf k}}) \biggr \}, \nonumber 
\end{align}
followed by relabeling of relevant terms, the Hamiltonian $H$ then takes the form, 
\begin{align}
\label{Ham3_1}
H^{(3)} &= \frac{1}{2}\biggl \{ \int \frac{1}{2}(V^{+++}_{kpq} + V^{+++}_{kqp})\cks \cp \cq \delkpq \krondelta  
 + (V^{++-}_{kpq} + V^{+-+}_{kqp}) \cks \cp \cqs \delkpq\krondelta
 \dkpq \biggr \}  \nonumber \\ & \quad \quad \quad + c.c..
\end{align}
The second term, $(V^{++-}_{kpq} + V^{+-+}_{kqp}) = \frac{{\bf p}^\perp \cdot {\bf q}_\perp}{k_\perp p_\perp q_\perp}(q_\perp - p_\perp)(k_\perp - p_\perp - q_\perp)$, evaluates to zero due to the delta function $\delkpq$. Hence we get the requisite Hamiltonian as follows:
\begin{equation}
H^{(3)} =  \int \biggl[ \tilde{L}_{kpq} \cks \cp \cq + c.c. \biggr]\delkpq \krondelta\dkpq
 \label{H3}
\end{equation}
where the interaction coefficient, $\tilde{L}_{kpq} := \frac{1}{4}(V^{+++}_{kpq} + V^{+++}_{kqp}) = \frac{1}{4} \frac{{\bf p}^\perp \cdot {\bf q}_\perp}{k_\perp p_\perp q_\perp}(q_\perp - p_\perp)(p_\perp + q_\perp) = \frac{1}{4} \frac{{\bf p}^\perp \cdot {\bf q}_\perp}{p_\perp q_\perp}(q_\perp - p_\perp)$ is symmetric in the second and third arguments, i.e. $\tilde{L}_{kpq} = \tilde{L}_{kqp}$. Here, the last equality is again due to the delta function $\delkpq$. Using the definition of total derivative, $\frac{\delta H^{(3)}}{\delta \cks} = \frac{\partial H^{(3)}}{\partial \cks} + \frac{\partial H^{(3)}}{\partial \cps}\frac{\delta \cps}{\delta \cks} + \frac{\partial H^{(3)}}{\partial \cqs}\frac{\delta \cqs}{\delta \cks}$ and the limit, $\delta \cks \to 0$, it is easy to show that Hamilton's equation \eqref{Ham_eqn_0} is satisfied,
where
\begin{align}
\frac{\delta H^{}}{\delta \cks} & =  \int \biggl[  \tilde{L}_{kpq} \cp \cq \delkpq \krondelta + 2\tilde{L}^*_{qpk} \cps \cq \delqpk \krondelta\biggr] \dpq. \label{dH3cks}
\end{align}
It must be noted here that equations \eqref{wave_amplitude1} and \eqref{Ham_eqn_0} are equivalent. 

\subsection{Wave Kinetic Equation}

By taking ensemble averages, i.e. $\langle\ck c^*_{\bf k'}\rangle$ in the above equation, we obtain an evolution equation for $e_{\bf k}\equiv e^+_{\bf k}=e^-_{\bf k}$ 
\begin{align}
\partial_\tau e_{\bf k} &= \Im \Biggl \{\int \Ls_{kpq}J_{kpq}\delkpq \krondelta -2\Ls_{pkq} J_{pkq}\delpkq \krondelta
\dpq \Biggr \}.\label{ek_unclosed}
\end{align} 
Here, $\Im (\cdot \cdot \cdot)$ refers to imaginary part of the argument in parenthesis, and  
\begin{equation}
\langle{\cks} \cp \cq\rangle := J_{kpq}(\tau) \delta (\Delta_{k,pq}),
\end{equation}
where $\Delta_{k,pq} := ({\bf k} - {\bf p} - {\bf q})$.\footnote{See footnote in sec.~\ref{tensor} for an explanation of the ensemble averaging.} The second term on the right hand side of equation (\ref{ek_unclosed}) is a consequence of simple algebraic manipulation of the second term on the right hand side of equation (\ref{dH3cks}).
We observe that equation~(\ref{ek_unclosed}) is not \textit{closed} in the sense that the left hand side of the equation is a second order correlation function that is expressed in terms of third order correlation functions on the right hand side. 

\subsubsection{Closure problem}
Here, we summarize the closure argument that may be generally applied to  wave-kinetic equations in  Hamiltonian form. On applying Wick's theorem to the Gaussian distributed wave field, quadruple correlation functions are defined as follows: 
\begin{equation}
\langle \cks \cps \cq \cm \rangle:= 2e_{\bf k}e_{\bf p}\delta(\Delta_{{{ k}{p},{q}{m}}}),
\label{4-correlator}
\end{equation}
where $\Delta_{{{ k}{ p},{ q}{ m}}} := ({\bf k} + {\bf p} - {\bf q} - {\bf m})$.\footnote{See footnote in sec.~\ref{tensor} for an explanation of the ensemble averaging.} Using this along with the definition of $J_{kpq}$ and equation~(\ref{wave_amplitude1}), it is possible to write a simple ordinary differential equation for $J_{kpq}$. Recall, $\partial_t \to \partial_t + \epsilon \partial_\tau$.\footnote{The total derivative is given by $\frac{d}{dt} = \frac{d}{dt} + \frac{d}{d\tau}\frac{d\tau}{dt}$ and since $\tau = \epsilon t$, we have $\frac{d\tau}{dt} = \epsilon$.} Allowing only triadic wave interactions, we begin by taking a fast time ($t$) derivative of $J_{kpq}(\tau)e^{i\phi(\omega) t} = J_{kpq}(\tau)e^{i\bigl (\phi(\omega) +  \phi^\prime(\omega)\bigr)t} = J_{kpq}(\tau)e^{i\bigl (\phi(\omega) + \epsilon \phi(\omega)\bigr)t}$:
\begin{align}
i\partial_t (J_{kpq}(\tau)e^{i\phi(\omega) t}) & \to -\phi(\omega)J_{kpq}(\tau)e^{i\phi(\omega) t} + \epsilon ie^{i\phi(\omega) t}\partial_\tau \langle \cks \cp \cq\rangle.
\label{J1_ODE}
\end{align}
Here $\phi(\omega) = (\omega_{\bf k} - \omega_{\bf p} - \omega_{\bf q}) \sim \mathcal{O}(1)$ and $\phi^\prime(\omega) \sim \mathcal{O}(\epsilon)$\footnote{Recall $\omega_k = \frac{k_Z}{k_\perp}$, and $\epsilon \sim \frac{k_Z}{k_\perp}$ is the small parameter.}  is the phase fluctuation. Consequently, we obtain the following relation at $\mathcal{O}(\epsilon)$:
\begin{align}
0 &= -\phi(\omega)J_{kpq}(\tau)e^{i\phi(\omega) t} + ie^{i\phi(\omega) t}\partial_\tau \langle \cks \cp \cq\rangle.
\label{J_ODE}
\end{align}
On applying the product rule for the second term  on the right hand side and using eqs.~\eqref{wave_amplitude1} and \eqref{4-correlator}, we obtain the relation:
\begin{align}
0 &= -\phi(\omega)J_{kpq} + {C_0}{} {\Ls}_{kpq}(e_{\bf k}e_{\bf p} + e_{\bf k}e_{\bf q} - e_{\bf p}e_{\bf q}), \label{ODE}
\end{align}
where $C_0$ is a dimensional constant with dimensions of $\frac{1}{L}$. The dimensions of $\delta(k)$ is $\frac{1}{k}$, $\krondelta$ is dimensionless and that of $dp dq$ is $k^2$, where $k \sim \frac{1}{L}, \omega \sim \frac{1}{T}$.\footnote{The fundamental dimensions of mass, length and time are denoted by $M, L, T$.} The second term on the right hand side is a direct outcome of using Wick's theorem on the terms that emerge after applying the product rule to the slow time derivatives. The applicability of Wick's theorem relies on the assumption that the field is Gaussian distributed and this point is further discussed in sec.~\ref{Comp_WT_Newell}. Note that $\phi(\omega) \equiv (\omega_{\bf k} - \omega_{\bf p} - \omega_{\bf q})$.

Thus, we have
\begin{equation}
J_{kpq}(\tau) = \frac{{C_0}}{\phi(\omega)}{\Ls^{}_{kpq}(e_{\bf k}e_{\bf p} + e_{\bf k}e_{\bf q} - e_{\bf p}e_{\bf q})}.
\end{equation}
The occurrence of the singularity ($\phi(\omega) = 0$) is averted by circumventing the pole by adding a term $i{\delta}$ (s.t. $\delta^2 \sim \epsilon \ll 1$) to the denominator and then multiplying the numerator and denominator terms by the complex conjugate of the denominator. Then we use the identity: $\Im\{ \delta_\omega + i{\delta}\} = -\pi {\delta}(\delta_\omega)$ and substitute the resulting term for $J_{kpq}$ in eq.~\eqref{ek_unclosed} to obtain the closed form of the three-wave kinetic equation:
\begin{align}
 \partial_\tau e_{\bf k} 
&= \frac{\pi C_0}{2}\int \Bigl [  |\Ls_{kpq}|^2 ( e_{\bf p} e_{\bf q} - e_{\bf k} e_{\bf p} -e_{\bf k} e_{\bf q})\delkpq \delta_{\omega^{}_k, \omega^{}_p  \omega^{}_q}\nonumber \\ &\qquad +2 |\Ls_{pkq}|^2 (e_{\bf k}e_{\bf q} - e_{\bf p}e_{\bf k} - e_{\bf p}e_{\bf q})\delpkq \delta_{\omega^{}_p, \omega^{}_k  \omega^{}_q}\Bigr ] \dpq. \label{ek_closed}
\end{align} 
Note that the occurrence of the product $\krondelta \delta_{\omega^{}_k, \omega^{}_p  \omega^{}_q}$ in the integrand implies that only $\delta_{\omega^{}_k, \omega^{}_p  \omega^{}_q}$ suffices as the frequency resonance condition as stated in eq.~(\ref{ek_closed}). Equation~(\ref{ek_closed}), that describes the evolution of the energy spectral density, can be rewritten as a single integral with the interaction operator proportional to the square of $\Ls_{kpq}$ by using the Zakharov-Kuznetsov conformal transformation as is explained in \citet{Zakha92,Kuznet72}. 

\subsubsection{Dimensional consistency of the wave kinetic system}
In this section, we verify that the derived equations are dimensionally consistent. To this effect it will suffice to show that eqs.~\eqref{wave_amplitude1} and \eqref{ek_closed} are dimensionally consistent. We will use the notation $[ \zeta ]$ to denote the dimension of the quantity $\zeta$. Recall, $[E_k] = \frac{L^3}{T^2}$, $[e_k]=\frac{L^4}{T^2}$, $[c_k] = [W_k] = \frac{L^2}{T}$. It is easy to check that $[c_k c^*_k] = [e_k]$ as is expected from the definitions in the paragraph following eq.~\eqref{U_tensor1}. 

Let us begin by checking the dimensionality of the l.h.s. and r.h.s. of eq.~\eqref{wave_amplitude1}. $[l.h.s.] = [\frac{c_k}{\tau}] = \frac{L^2}{T^2}$. Now, $[r.h.s.] = [V_{kpq}][c^2][\delkpq][\krondelta][dp dq] = \frac{1}{L}\frac{L^4}{T^2}L\frac{1}{L^2} = \frac{L^2}{T^2}$. Therefore, eq.~\eqref{wave_amplitude1} is dimensionally consistent. 

Now let us verify if eq.~\eqref{ek_closed} is dimensionally consistent. Like before, $[l.h.s.] = [\frac{e_k}{\tau}] = \frac{L^4}{T^3}$ and $[r.h.s] = [C_0][\Ls^2_{kpq}][e^2][\delkpq][\delokpq][dp dq] = \frac{1}{L}\frac{1}{L^2}\frac{L^8}{T^4}LT\frac{1}{L^2} = \frac{L^4}{T^3}$. Therefore, the derived kinetic equation is dimensionally consistent. 

\subsubsection{Invariants of the closed three-wave kinetic equation}
It can readily be  shown that the total energy given by equation~(\ref{ek_closed}) is conserved, i.e. $\partial_t \int e_{\bf k} d{\bf k} = 0$ using the result of \textbf{Appendix \ref{invariance_proof}}. The proof in the appendix is essentially a statement of conservation of energy in each wave triad. The squared interaction operator, $|\Ls_{kpq}|^2 \sim \frac{1}{k^2_{\perp}}$ ensures the convergence of the collision integral appearing on the r.h.s. of equation~(\ref{ek_closed}), thereby it meets the first criteria for the realizability of a Kolmogorov spectrum, i.e., stationary energy spectrum solution to (\ref{ek_closed}) \citep{Balk90}. While possible, a stability analysis of the spectra for the anisotropic medium is beyond the scope of this work and the interested reader may be referred to \citet{Balk90}. 

\subsubsection{Kolmogorov Solution of the Three-Wave Kinetic Equation}
\label{Kol}
On assuming the  closure argument provided above, the exact solution of the three-wave kinetic equation \eqref{ek_closed}, as power laws, is obtained by assuming locality of the scale-by-scale energy transfer. This is illustrated in \citet{Zakha92} and earlier papers referenced therein.

The four possible stationary solutions for the anisotropic spectrum, $e_{\bf k} \sim k^{-x_i}_Z k^{-y_i}_{\perp}, \forall k_z \ne 0$, are listed as follows:
\begin{enumerate}
\item[(i)] $x_1 = 1$ and $y_1 = -1$.
\item[(ii)] $x_2 = 1$ and $y_2 = 0$.
\item[(iii)] $x_3 = (1 + u)$ and $y_3 = (2 + v)$, where $2u$ and $2v$ are respectively  the powers of $k_z$ and $k_{\perp}$ in $|\Ls_{kpq}|^2$. Clearly, in our case, $u=0,v=1$ (equation~(\ref{H3})). Thus, $x_3 = 1$ and $y_3 = 3$.
\item[(iv)] $x_4 = 1$ and $y_4 = 7/2$. This solution corresponds to the constant flux in the z-component of the momentum.
\end{enumerate}
Solution (iii), above, corresponds to the \textit{only} constant energy flux solution, a necessary requirement of  Kolmogorov's theory and of primary concern in this paper. The constancy of energy flux can be easily verified by noting that $\partial_t E_{\bf k} := \partial_t (2\pi k_{\perp} e_{\bf k}) = -\partial_{k_{\perp}}\Pi(k_{\perp},k_Z;t)$; whereby on integrating with respect to $k_{\perp}$ and demanding constant energy flux in the perpendicular direction, we can extract the aforementioned solution. Here, $\Pi$ denotes the flux of energy. 
Thus, the exact solution for the Kolmogorov-Zakharov-Kuznetsov spectra with constant flux is as follows:
\begin{equation}
 e_{\bf k} = e(k_{\perp},k_Z) \sim k_{\perp}^{-3}k_{Z}^{-1}.
 \label{exact_ek_spec}
\end{equation} 
This result is in agreement with experimental and computational simulations\citep{Thiele09, Mininni12, Teitel12}.\\
 \section{Non-zero helicity dynamics: interplay of energy and helicity}
 In this section, we deduce a general set of coupled equations for the two invariants of the flow. This is done by formally extending the symmetrical system of the previous section.
 \subsection{Coupled equations for energy and helicity}
 Recall that the assumption, $c^-_{\bf k} = c^*_{\bf k} \equiv c^{+*}_{\bf k}$  implies $h_{\bf k} = 0$, i.e. $\frac{1}{2}e_{\bf k} = e^+_{\bf k} = e^-_{\bf k}$. However, on relaxing such an assumption, a coupled set of equations for energy and helicity may be arrived at by using equations \eqref{eh_coupled_1} in equation \eqref{ek_closed} (\textbf{cf.} $e_{\bf k}$ in equation \eqref{ek_closed} is actually $e^+_{\bf k} \equiv e^-_{\bf k}$). Thus the closed form coupled energy-helicity equation becomes,  
 \begin{align}
 \partial_\tau \biggl(e_{\bf k} \pm \frac{h_{\bf k}}{k_\perp}\biggr)&= \frac{\pi C_0}{4}\int \Biggl [  |\Ls_{kpq}|^2 \biggl \{ \biggl(e_{\bf p} \pm \frac{h_{\bf p}}{p_\perp}\biggr) \biggl(e_{\bf q} \pm \frac{h_{\bf q}}{q_\perp}\biggr)- \biggl(e_{\bf k} \pm \frac{h_{\bf k}}{k_\perp}\biggr) \biggl(e_{\bf p} \pm \frac{h_{\bf p}}{p_\perp}\biggr) \nonumber \\ &\quad - \biggl(e_{\bf k} \pm \frac{h_{\bf k}}{k_\perp}\biggr) \biggl(e_{\bf q} \pm \frac{h_{\bf q}}{q_\perp}\biggr) \biggr \}\delkpq \delta_{\omega^{}_k, \omega^{}_p  \omega^{}_q} + 2 |\Ls_{pkq}|^2 \biggl \{ \biggl(e_{\bf k} \pm \frac{h_{\bf k}}{k_\perp}\biggr)\biggl(e_{\bf q} \pm \frac{h_{\bf q}}{q_\perp}\biggr) \nonumber \\ & \quad \quad - \biggl(e_{\bf p} \pm \frac{h_{\bf p}}{p_\perp}\biggr)\biggl(e_{\bf k} \pm \frac{h_{\bf k}}{k_\perp}\biggr)- \biggl(e_{\bf p} \pm \frac{h_{\bf p}}{p_\perp}\biggr)\biggl(e_{\bf q} \pm \frac{h_{\bf q}}{q_\perp}\biggr) \biggr \} \delpkq \delta_{\omega^{}_p, \omega^{}_k  \omega^{}_q} \Biggr ] \dpq. \label{ek_closed_coupled}
 \end{align} 
The individual evolution equation for $e_{\bf k}$ (and $h_{\bf k}$) follows by adding (and subtracting) the two set of equations expressed concisely by equation \eqref{ek_closed_coupled} and is given as follows:
\begin{align}
 & \partial_\tau e_{\bf k} \nonumber \\
&= \frac{\pi C_0}{4}\int \Biggl [  |\Ls_{kpq}|^2 \biggl \{ \biggl( e_{\bf p} e_{\bf q} + \frac{h_{\bf p} h_{\bf q}}{p_\perp q_\perp}\biggr) - \biggl( e_{\bf k} e_{\bf p} + \frac{h_{\bf k} h_{\bf p}}{k_\perp p_\perp}\biggr) - \biggl( e_{\bf k} e_{\bf q} + \frac{h_{\bf k} h_{\bf q}}{k_\perp q_\perp}\biggr)\biggr \} \delkpq \delta_{\omega^{}_k, \omega^{}_p  \omega^{}_q}   \nonumber \\ 
& \quad + 2 |\Ls_{pkq}|^2 \biggl \{ \biggl( e_{\bf k} e_{\bf q} + \frac{h_{\bf k} h_{\bf q}}{k_\perp q_\perp}\biggr) - \biggl( e_{\bf k} e_{\bf p} + \frac{h_{\bf k} h_{\bf p}}{k_\perp p_\perp}\biggr) - \biggl( e_{\bf p} e_{\bf q} + \frac{h_{\bf p} h_{\bf q}}{p_\perp q_\perp}\biggr)\biggr \}  \delpkq \delta_{\omega^{}_p, \omega^{}_k  \omega^{}_q} \Biggr ] \dpq, \label{ek_coupled}
\end{align}
and
\begin{align}
 &\partial_\tau h_{\bf k}  \nonumber \\
&= \frac{\pi C_0}{4}\int k_\perp \Biggl [  |\Ls_{kpq}|^2 \biggl \{ \biggl( e_{\bf p} \frac{h_{\bf q}}{q_\perp} + e_{\bf q}\frac{h_{\bf q}}{q_\perp}\biggr) - \biggl( e_{\bf k} \frac{h_{\bf p}}{p_\perp} + e_{\bf p}\frac{h_{\bf k}}{k_\perp}\biggr) - \biggl( e_{\bf k} \frac{h_{\bf q}}{q_\perp} + e_{\bf q}\frac{h_{\bf k}}{k_\perp}\biggr)\biggr \} \delkpq \delta_{\omega^{}_k, \omega^{}_p  \omega^{}_q}   \nonumber \\ 
& \quad + 2 |\Ls_{pkq}|^2 \biggl \{ \biggl( e_{\bf k} \frac{h_{\bf q}}{q_\perp} + e_{\bf q}\frac{h_{\bf k}}{k_\perp}\biggr) - \biggl( e_{\bf k} \frac{h_{\bf p}}{p_\perp} + e_{\bf p}\frac{h_{\bf k}}{k_\perp}\biggr)  - \biggl( e_{\bf p} \frac{h_{\bf q}}{q_\perp} + e_{\bf q}\frac{h_{\bf q}}{q_\perp}\biggr) \biggr \}  \delpkq \delta_{\omega^{}_p, \omega^{}_k  \omega^{}_q} \Biggr ] \dpq. \label{hk_coupled}
\end{align}
Equations \eqref{ek_coupled} and \eqref{hk_coupled} clearly reveal the coupled dynamics of energy and helicity in a rapidly rotating fluid flow. The possibility of recovering the total energy and helicity spectrum from the simpler zero helicity case relies on extending the functional on the right hand side of equation~(\ref{ek_closed}) to the non-zero helicity case. 

In essence, we have taken a special case of the wave-kinetic equations  which has the functional form $\partial_t e_{\bf k} = f({\bf k})$ (think of $f({\bf k})$ as the right hand side of equation \eqref{ek_closed}) where $h_{\bf k} = 0$, and extended it to the more general case where $h_{\bf k} \ne 0$. In this general case, the domain of $f({\bf k})$ is still the positive real line because the inequality $|h_{\bf k}| \le k_\perp e_{\bf k}$ implies that $h_{\bf k}$ and $e_{\bf k}$ are not independent variables and $e_{\bf k} \pm \frac{h_{\bf k}}{k_\perp} \ge 0$ is always true. Hence, we have extended the applicability of the wave kinetic equation from the simpler symmetric case (where $c^*_{\bf k} = c^{-}_{{\bf k}}$ because $h_{\bf k} = 0$) to the more general case with non-trivial helicity. This way we have circumvented tedious algebraic computations involving multiple correlation functions (to account for the departure in mirror symmetry in the fully helical case) by first, reducing the Hamiltonian system by invoking \textit{reflection symmetry} and then formally extending the symmetrical system to the more general helical case. The applicability and implication of this procedure is explained in more details in \citep{Sen_th}.

\subsection{Generalized solution of energy and helicity spectra}
Note that eqs. \eqref{eh_coupled_1} are consistent with the definitions, $e_{\bf k} = {e^+_{\bf k} + e^-_{\bf k}}{}$ and $h_{\bf k} = k_{\perp}({e^+_{\bf k} - e^-_{\bf k}}{})$. Clearly, the power law solutions do not change for the non-zero helicity case because the homogeneity of the interaction coefficient and the linear dispersion relation remains the same. Thus, $e_{\bf k} \sim k^{-3}_{\perp}$ for $h_{\bf k} \ne 0$. 
Dimensional analysis implies, $h_{\bf k} \sim k_\perp e_{\bf k} \sim  k^{-2}_\perp$ that is consistent with earlier findings based on numerical simulations\citep{MinPou10}. The cylindrically symmetric solutions are:
\begin{equation}
E_{k_\perp} = (2\pi k_\perp e_{k}) \sim k^{-2}_{\perp}, \text{ and } H_{k_\perp} = (2\pi k_\perp h_{k}) \sim k^{-1}_{\perp}. \label{gen_E_H_soln}
\end{equation}
The power law solution obtained here pertains to the perpendicular cascade within the slow manifold, the reader is referred to \citet{Bellet06} for power law solutions of the spectrum in the axial direction. It is important to note that the solutions given by equation \eqref{gen_E_H_soln} are different from the ones discussed in \citet{Pouq10} as the latter assume {isotropy} in their arguments to show that the sum of the powers of the wave numbers for $E_{\bf k}$ and $H_{\bf k}$ is equal to 4. The solution set of \citet{Galtier03} belongs to the regime where the effect of fast inertial waves is dominant. In the recent work of \citet{Galtier14}, it has been shown that within the theory developed in the earlier work of \citet{Galtier03}, the power law solutions can be generalized to the empirical form presented in \citep{Pouq10}. In contrast, the analysis presented here captures the dynamics in the slow manifold where the flow is highly anisotropic and the effect of the fast inertial waves is sub-dominant, and accounts for a distinct separation of temporal scales, $t$ and $\tau$ as is necessary \citep{EM98} and explained in the introductory section here. In such a regime in the slow manifold that is devoid of fast inertial wave modulation and where the dynamic is well captured by the R-RHD equations, the stationary power law solutions, that are entirely dependent on the functional form of the interaction term $\Ls$ and the dispersion relation $\omega_{\bf k}$, are given by eq.~(\ref{gen_E_H_soln}) and is in agreement with results from numerical simulations (e.g., the work of \citet{MinPou10}, \citet{Teitel12}). 

The important point to note here is that rotating turbulence encompasses several different and distinct dynamical regimes, each with distinct set of solutions. Hence, the importance of using reduced equations for distinct asymptotic limits as has been explained by \citet{NS11} (see appendix A of \citep{NS11}). Within such a distinct limit of rapid rotation, the R-RHD equations have been derived by \citet{JKW98,kj06,Julien07,NS11} and a multiple scales perturbation method has been applied in this paper for the asymptotically reduced equations. In the following sections, we attempt to stitch together the important results of weak and strong turbulence of different dynamical regimes in order to clarify the picture of turbulent cascade that has evolved based on recent research literature. 

\subsection{Hierarchy of slow manifolds in anisotropic turbulence diverges from the critical balance route towards isotropy} 
The discussion in this section is motivated by the turbulence cascade picture presented by \citet{NS11}. We modify the turbulent cascade schematic based on the results presented here. 
 \begin{figure}[h!]
\begin{center}
\includegraphics[scale=0.5]{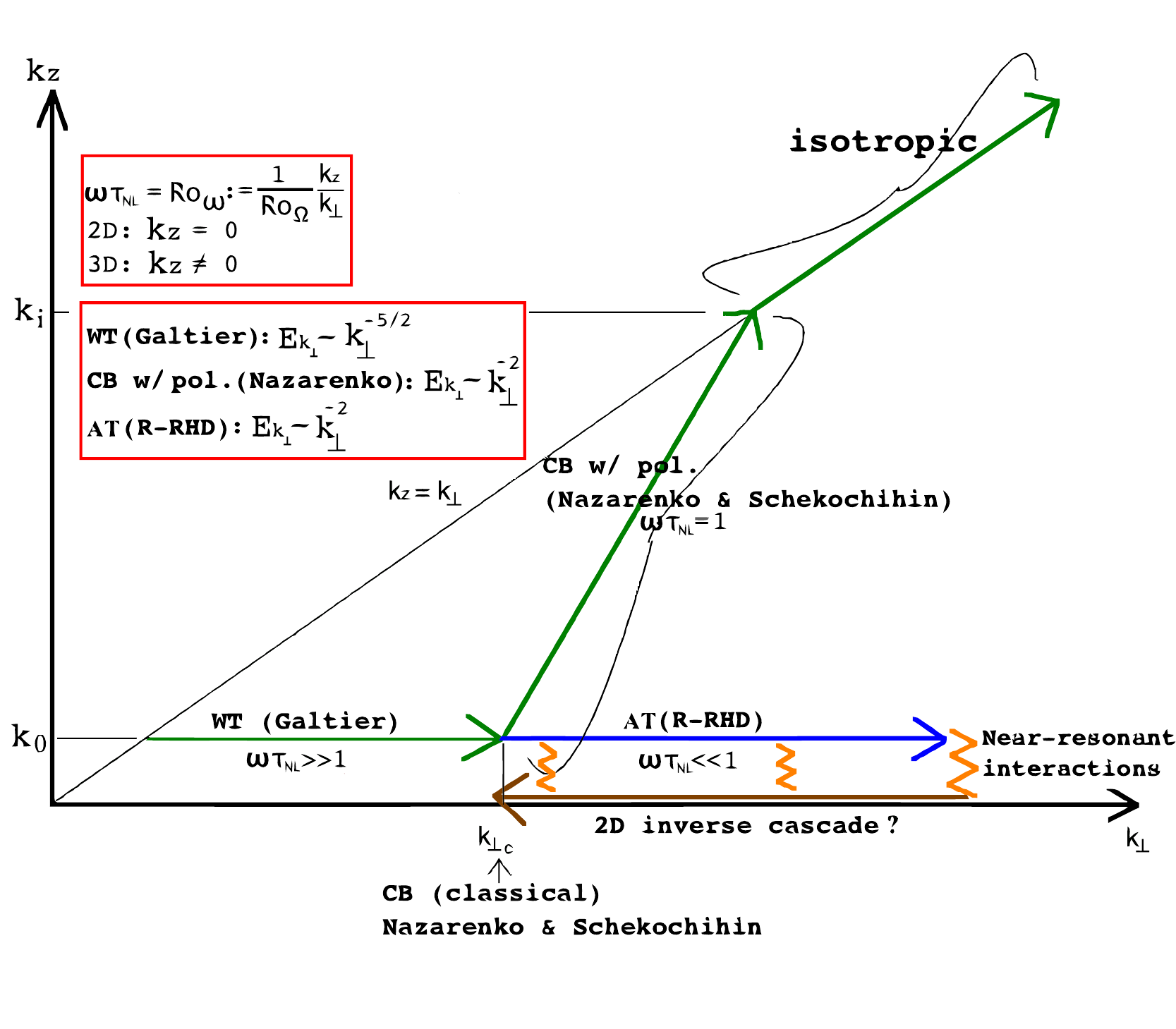}
\end{center}
\caption {\small{(Color online) A sketch of cascade paths for rotating turbulence shows the different flow regimes depending on $\mathcal{R}o_{\omega} (\equiv \omega \tau_{NL})$ and the corresponding energy spectra. Here, $k_i$ is the isotropic wavenumber, $k_{\perp c}$ is the classical critical balance wavenumber, $k_0$ is the injection wavenumber corresponding to an initial wave field. Three distinct regimes are shown: (i) WT (Galtier) corresponding to the wave-turbulence regime with $\mathcal{R}o_{\omega} \gg 1$, (ii) CB w/ pol. (i.e. critical balance with polarization alignment) as explained in \citep{NS11} leading towards isotropy, and (iii) AT (R-RHD) corresponding to the anisotropic turbulent dynamics of the R-RHD equations with $\mathcal{R}o_{\omega} \ll 1$. As we move along the horizontal axis from left to right, the flow traverses a hierarchy of slow manifolds with successively rescaled (decreasing) $k_Z/k_\perp$ wave number ratio. Also shown are possible explanation of 2D-3D coupling by non-resonant interactions. The AT (R-RHD) theory does not explain inverse cascade phenomena. Here WT stands for wave-turbulence, CB stands for critical balance and AT stands for anisotropic turbulence.}}
\label{WT_CB}
\end{figure}
The critical balance with polarization alignment argument presented by \citet{NS11} is an attempt to explain the $k^{-2}_\perp$ energy spectra observed in several numerical simulations of rotating turbulence \citep{Thiele09, MinPou10, Teitel12,TMin09}. However, as is evident from figure \ref{WT_CB} and the corresponding sketch in \citep{NS11}, the critical balance with polarization alignment leads to a departure from anisotropy and is a path to the recovery of isotropic scales that are prevalent above the Zeeman wavenumber \citep{Mininni12}. It must be emphasized that anisotropy is dominant in rotating turbulent flows and this is verified by several numerical simulations and experiments that have been cited in this paper. It is in this light, we believe that the energy spectra $~ k^{-2}_\perp$ is obtained for anisotropic turbulence derivation in the regime ($\omega \tau_{NL} \equiv \mathcal{R}o_{\omega} := \frac{1}{\mathcal{R}o_{\Omega}}\frac{k_z}{k_\perp} \sim k_Z/k_\perp \ll 1$) of ~R-RHD. This is further corroborated by numerical simulations where the $k^{-2}_\perp$ energy spectra corresponds to $\mathcal{R}o_{\omega} \ll 1$  (see Table I in \citep{Teitel12}). In this anisotropic regime, the slow inertial wave frequency is much smaller than the fast inertial wave frequency. This solution prevails as the flow traverses a hierarchy of slow manifold regions with successively decreasing $\frac{k_Z}{k_\perp}$ and is the anisotropic turbulence solution for the energy spectra within the slow manifold. In summary, there seems to be a bifurcation of the energy spectral solution at the critical balance wavenumber (see figure \ref{WT_CB}), two distinct spectra evolve, each with a $k^{-2}_\perp$ energy spectra: one leading towards the isotropic scale with a micro-Rossby number $\mathcal{R}o_{\omega} = 1$ via critical balance and hence falls within the realm of strong turbulence, and the other towards highly anisotropic horizontal scales with $\mathcal{R}o_{\omega} \ll 1$ sustained by small amplitude resonating wave modes.  

\subsection{Comparison with weak turbulence theory of Galtier}
In this section, we contrast the theory presented here with the recent works of \citet{Galtier03, Galtier14}. The main distinctive features are listed below.
\begin{enumerate}
\item The governing equations on which the wave turbulence theory of \citet{Galtier03, Galtier14} is developed are the Navier-Stokes equations (i.e. eq.~(\ref{NSE})). In these equations, the Rossby number $\mathcal{R}o$ appears explicitly and is used as an order parameter in the perturbation analysis and the fast inertial wave dynamic is dominant. In contrast, the governing equations for the analysis presented here are the R-RHD (i.e. eqs.~(\ref{R-RHD1a}) and (\ref{R-RHD1b})). The asymptotic limit of infinitesimally small $\mathcal{R}o$ is already accounted for in the multiple scales analysis to derive the R-RHD and hence do not explicitly appear in the R-RHD equations. This means that the effect of fast inertial waves is sub-dominant here and the R-RHD equations are hence suitably applicable for the slow manifold dynamics. 
\item The dynamical regime where the theory of \citet{Galtier03, Galtier14} is valid is $\omega \tau_{NL} \gg 1$. This point has been elaborated in great detail in the work of \citet{NS11} (see sec. 2 in \citep{NS11}). The dynamical regime of the R-RHD is $\omega \tau_{NL} \ll 1$. It is the limit in which $k_z$ is so small that the turbulent dynamic is populated by slow inertial waves with dispersion relation $\omega_k = \frac{k_Z}{k_\perp}$ embodying slow oscillations because $\frac{k_Z}{k_\perp} \ll 1$ (also see FIG. \ref{WT_CB} above). In other words, within the slow manifold $k_z$ is so small that $\frac{k_z}{k_\perp} \ll \mathcal{R}o \ll 1$ and consequently the relation $\omega \tau_{NL} = \frac{1}{\mathcal{R}o}\frac{k_z}{k_\perp} \sim k_Z/k_\perp$ entails $\omega \tau_{NL} \ll 1$. This means $\omega \tau = \frac{\tau}{t} \ll 1$ or $\tau \ll t$. This led to the choice $\tau = \epsilon t$, where $\epsilon \sim k_Z/k_\perp \ll 1$. The perturbation analysis presented here is applied to the R-RHD in this dynamical regime where the smallness (weakness) of the wave amplitude is measured by $\epsilon \ll 1$ and the slow dispersive three wave system undergoes weak non-linear exchanges at the asymptotic order $\epsilon^2$ as explained earlier in sections \ref{WAMP} and \ref{ZeroHel}.
\item In the theory of \citet{Galtier03, Galtier14}, since the fast inertial waves are dominant, the small amplitudes evolve at the wave time scale $t$ (i.e. compare the terms $\partial_t a^s_{\bf k}$ and $e^{-i(s\omega_k - s_p\omega_p - s_q\omega_q)t}$ in eq.~(3) of \citet{Galtier03}). This means that the advective time scale $\tau$ does not naturally appear in the analytical derivation. In contrast, in the theory developed here, the fast inertial wave time scales are sub-dominant and are filtered out during the derivation of the R-RHD equations and only \textit{slow} inertial quantities are retained with $\omega_{\bf k} = \frac{k_Z}{k_\perp}$ where $k_Z = \frac{1}{\mathcal{R}o}k_z$ represents \textit{slow} oscillatory modes. Moreover, a distinct temporal scale separation of advective ($\tau$) and inertial ($t$) quantities is included in the multiple scales perturbation theory developed here for the resonant wave dynamics. This is in agreement with hypothesis (i) mentioned here in the introductory section and elaborated in the work of \citet{EM98}.  
\item In the theory of \citet{Galtier03, Galtier14}, the kinetic equations for energy and helicity are derived by using multiple  correlation functions to capture the energetics as well as the absence of symmetry due to helicity. This makes the calculations tediously lengthy. In contrast, the derivation for the helicity kinetic equation is presented here as a natural extension of the symmetrical non-helical system and bypasses the use of calculations using multiple correlation functions. This simpler approach follows a more general philosophy of \textit{Hamiltonian reduction} exploiting symmetries in the system \citep{TS90} and their natural extension to understanding asymmetrical phenomena. 
\end{enumerate}
Despite the fundamental differences in the region of validity of the two theories, they present a more detailed recipe to better understand turbulent energetics and cascades for rotating flows. A simple schematic towards this goal is presented in FIG. \ref{WT_CB} above to highlight the key findings in this field.  

\subsection{Comparison with weak-wave turbulence theory: cumulant hierarchy vs wave amplitude hierarchy}
\label{Comp_WT_Newell}
The fundamental difference between the multiple scales perturbation technique employed in this paper to derive the kinetic equations and the weak wave turbulence theory reviewed by Newell et. al \citep{Newell11} is explained in the following paragraphs. 

Weak turbulence calculations, as reviewed by Newell et. al. \citep{Newell11}, are computed in the regime $\tau_{NL} \gg t$, the closure is obtained by further taking the limit $t \to \infty$. The third and higher order cumulants survive at the longer advective timescale ($\tau_{NL}$). This implies prevalence of non-Gaussian statistics for long advective time span. This is the reason for writing a hierarchal system for the cumulants in the perturbation approach, {\bf cf.} Newell et. al. \citep{Newell11}, rather than for the Fourier amplitudes.  

On the contrary, the theory developed in this paper belongs to the opposite regime, i.e. $\tau_{NL} \ll t$. The dynamics explained here elapse at the much slower advective time scale ($\tau_{NL}$) compared to the faster wave timescale $t \to \infty$. At this slower time scale, third (and higher odd) order moments (cumulants) are sub-dominant thereby corroborating the assumption of Gaussian statistics for the wave field. This justifies the applicability of Wick's theorem in the derivation of the kinetic equations. Consequently, this is also why we employed a hierarchal system for the wave amplitude, {\bf cf.} eq.~\eqref{amp_expansion}, rather than expanding the cumulants with higher order correction terms. In the regime $\omega \tau_{NL} \ll 1$, the higher order corrections are not required and hence absent. 

\section{Coupling between wave and 2D modes through non-resonant interactions}
Note that $\Ls_{kpq} = 0$ when $k_Z=0$ because of the fact that $p_\perp = q_\perp$. Thus, within the framework of purely resonating wave triads, it is not possible to establish the coupling between wave and purely 2D modes. This should not be surprising because the theory developed is that of dispersive waves that are in resonance. However, the coupling with 2D modes can be explained by a small modification, as is presented below. 

Suppose that in equation \eqref{wave_amplitude1}, the triadic resonance condition is modified such that $\phi(\omega) = s_k\omega_k - s_p\omega_p -s_q\omega_q = \delta_\omega \ll \epsilon, \delta_\omega \ne 0$. This condition represents non-resonant (or near-resonant) interaction of the three wave modes. 
For small $\frac{\delta_\omega}{\epsilon}\tau$, Taylor expansion implies,
\begin{equation}
e^{i \frac{\delta_\omega}{\epsilon}\tau}   \approx 1+ i\frac{\delta_\omega}{\epsilon}\tau + ...
\end{equation} 
i.e.,
\begin{equation}
\label{non_res1}
e^{i \frac{\delta_\omega}{\epsilon}\tau} \approx  \biggl( \delta_{\omega^{s_k}_k, \omega^{s_p}_p  \omega^{s_q}_q}\biggr )_{\delta_\omega=0} +i\frac{\delta_\omega}{\epsilon}\tau + ...~.
\end{equation}
Equation \eqref{wave_amplitude1} is arrived at after computing an inner-product in wavenumber and time, the latter converts an exponential term, $e^{i(s_k \omega_k - s_p \omega_p - s_q \omega_q)t}$, to the frequency delta function in eq. \eqref{wave_amplitude1}. However, computing the fast time integral after Taylor-expanding the exponential term reveals that the higher order \textit{slow} secular terms ({\bf cf.} presence of $\tau$ in the higher order terms) are inherently embedded in the full system that is not restricted to resonating triadic interactions only. These terms account for the coupling with the 2D modes. This is evident by re-writing eq.~\eqref{wave_amplitude1} with the higher order terms as follows,  
\begin{align}
\label{wave_amplitude_nonres1}
i\partial_\tau c^{s_k}_{\bf k} & = \frac{1}{2T}\sum_{s_p,s_q}\int V^{s_ks_ps_q}_{kpq}c^{s_p}_{\bf p}c^{s_q}_{\bf q}  \biggl[ \underbrace{{\biggl( \delta_{\omega^{s_k}_k, \omega^{s_p}_p  \omega^{s_q}_q}\biggr )_{\delta_\omega=0}}}_{\hbox{res}} + \underbrace{i\frac{\delta_\omega}{\epsilon}\tau + ...}_{\hbox{non-res}} \biggr] \delkpq \dpq dt.
\end{align}The kinetic equation, that is constructed from the above amplitude equation as explained before, can then be decomposed into two parts with contributions from resonating and non-resonating modes considered separately, as follows:
\begin{equation}
\label{non-res_tr}
\partial_\tau e_{\bf k} = T_{\text{res}}({\bf k},\tau) + T_{\text{non-res}}({\bf k},\tau),
\end{equation}
where $T({\bf k},\tau)$ denotes nonlinear transfer of energy to mode $\bf k$.
Thus, retaining only non-resonant interactions, it is possible to establish the aforementioned coupling in the slow manifold, $k_Z = 0$. Note that in the case of non-resonant interactions, due to the absence of the frequency delta function, a stationary Kolmogorov (constant flux) solution cannot be obtained. The reader is also referred to the works of \citet{Janssen03} and \citet{Annenkov06} for a detailed theory of quasi-resonant interactions in four wave dispersive systems. 

\section{Conclusion} 
In conclusion, it is important to emphasize an important point, that of the asymptotic dynamical regime to which the fluid system belongs. In the context of this paper, we have restricted our analysis to the highly anisotropic regime of \textit{rapid} rotation (i.e. infinitesimally small Rossby number) within the slow manifold where $k_z$ is infinitesimally small. It has been shown that the application of a rigorous multiple scales perturbation method within this regime yields a $k^{-3}_\perp$ law for the anisotropic energy spectrum, $e_{\bf k}$ ({\bf cf.} equivalently a $k^{-2}_\perp$ spectrum for the cylindrically symmetric spectrum, $E_{k_\perp}$) that is in agreement with results from numerical simulations as has been stated earlier. An asymptotically reduced system spans a hierarchy of slow manifold regimes and thereby captures the gradual transfer of energy towards the {2D} modes. Interestingly, a similar power law solution can also be obtained by applying a \textit{critical balance} phenomenology (where fast inertial wave time scale balances the nonlinear advection time scale) to the system of rotating turbulence as has been shown in \citet{NS11}. This is the realm of strong turbulence where the nonlinear interactions are strong, meaning $\omega \tau_{NL} \sim \mathcal{O}(1)$. However, the anisotropic limit of rapidly rotating turbulence is farther away from modes where critical balance holds, this has been shown through numerical simulations in \citet{Leoni13}. In addition to the discussion in section V(C) above, the reader is referred to \citep{Sen_th, NS11} for a detailed discussion on a wave turbulence and critical balance schematic of the energy cascade process. It is important to note that in the analysis presented in this paper, any physical artifact induced by boundary condition is not considered. Interested readers are referred to the work of \citep{Elena10} that describes discrete boundary effects on wave turbulence formalism. \\

In summary, we have constructed a statistical wave kinetic theory for an asymptotically reduced set of equations that is valid in the limit of rapid rotation and in the anisotropic slow manifold regime. Stationary solutions of invariant quantities have been obtained that are consistent with experimental and simulation data reported in recent work. A coupled set of equations has been derived explaining the nature of the inter-dynamics of the two global invariants of the system, viz., energy and helicity. This has been done by extending the \textit{symmetrical} non-helical system to the more general helical case where the reflection symmetry is broken. This procedure is novel in the sense that it bypasses construction of multiple correlation functions to account for the departure in mirror symmetry in the helical case. This analytical study will serve a useful reference point for theoretical understanding of atmospheric phenomena of planets that require a better knowledge of anisotropic wave dynamics.

\section{Acknowledgments}
The author would like to acknowledge useful comments by Pablo D. Mininni, Annick Pouquet and Keith Julien. The department of Applied Mathematics at the University of Colorado, Boulder is also acknowledged for providing financial support through graduate assistantship and travel grants to the author while he was a doctoral student. This paper is dedicated to Dr. Richard Juday and Dr. Darcy Juday for their constant inspiration and encouragement.

\section{Appendix}

\subsection{Natural appearance of the frequency delta function in the amplitude equation}\label{freq_delta_amp_eq}
The frequency delta function in wave amplitude eq.~(\ref{wave_amplitude1}) appears naturally on applying the solvability condition to eq.~(\ref{R-RHD_ham0}). This is explained in detail in this section. 

Recall from section \ref{WAVEFIELD} that the wave field is proportional to terms that evolve at advective time scale $\tau$ and the exponential term that elapses at the inertial time scale $t$ (\textbf{cf.} since the R-RHD limit is $\omega \tau = \frac{\tau}{t} \ll 1$, we chose $\tau = \epsilon t, \epsilon \ll 1$, also $t \ll $ fast inertial time scale that has been filtered out by the R-RHD),
\begin{equation}
\mathbf{\Psi}^{s_k}_{\bf k}e^{i{\bf \Phi}({\bf k}, s_k\omega_{\bf k}t)} \sim \underbrace{{\bf h}^{s_k}_{\bf k} c^{s_k}_{\bf k}(\tau)}_{\text{advective scale }\tau}\underbrace{e^{i{\bf \Phi}({\bf k}, s_k\omega_{\bf k}t)}}_{\text{inertial scale } t}.
\label{scale_sep}
\end{equation}  
This separation of scales between $\tau$ and $t$  (note $\tau$ and $t$ are now independent variables) allows us to {average (integrate) out the exponential term over the inertial timescale $t$} and leaves us with terms that are dependent on the advective scale $\tau$ alone. This enables us to write an evolution equation for the small amplitude of the form ${\partial_\tau c^{s_k}_{\bf k}(\tau) = \text{r.h.s of eq.~(\ref{wave_amplitude1})}}$. This is standard procedure in multi-scale perturbation techniques and the interested reader is referred to the comprehensive book on this topic by \citet{Bender99}. To elucidate that the averaging is done over a large time limit, consider eg. $\epsilon = \frac{1}{1000}$, consequently $\tau = \epsilon t = \frac{1}{1000}t$; this means that for $\tau$ to elapse 1 unit, $t$ must elapse 1000 units, i.e. within the scope of $\tau$, $t$ is already very large. The average of a function, say $f(t)$ defined over the domain $D$ is given as
\begin{equation}
f_{avg} = \frac{1}{\text{size of } D}\int_{D} f(t) dt = \frac{1}{2t}\int_{-\tlim}^{\tlim} f(t) dt. \label{averaging}
\end{equation}
We apply the solvability condition $\frac{1}{2}({\bf h}^{-s_k}_{\bf k} \cdot {\cal L_H}{\bf U}^{s_k}_{2\bf k})= 0$ to eq.~(\ref{R-RHD_ham0}) to obtain the amplitude equation. The solvability condition involves an inner product (denoted by the operator $\cdot$ above) that includes a projection onto the helical basis (that we denote by angle brackets here, i.e. $\langle \frac{1}{2}{\bf h}^{-s_k}_{\bf k}  , \cdot \rangle$ in wavenumber space) and a time averaging (that we denote by normal integration symbol) as shown below
\begin{equation}
\label{inner_prod}
\frac{1}{2t}\int_{-\tlim}^{\tlim} \langle \frac{1}{2}{\bf h}^{-s_k}_{\bf k} , \cdot \rangle dt. 
\end{equation}
This time integration is over the large inertial time limit, $t \sim \frac{1}{\omega_{\bf k}}$ where $\omega_{\bf k}$ is asymptotically small because $\frac{k_Z}{k_\perp} \ll 1$. We operate each term of eq.~(21) with the operator (\ref{inner_prod}) defined above.\\\\
\textbf{{1st term}}: $\frac{1}{2}({\bf h}^{-s_k}_{\bf k} \cdot {\cal L_H}{\bf U}^{s_k}_{2\bf k}) := \frac{1}{2t}\int_{-\tlim}^{\tlim} \langle \frac{1}{2}{\bf h}^{-s_k}_{\bf k} , {\cal L_H}{\bf U}^{s_k}_{2\bf k} \rangle dt  = \frac{1}{2t} \int_{-\tlim}^\tlim 0 dt = 0$. The $0$ integrand follows from application of the solvability criterion which is basically a \textit{Fredholm alternative} in the context of the adjoint problem explained in section \ref{WAMP}. This makes the left hand side of eq.~(\ref{R-RHD_ham0}) null upon application of the inner product.\\\\
\textbf{{2nd term}}: $\frac{1}{2t}\int_{-\tlim}^\tlim \partial_\tau c_{\bf k}^{s_k}(\tau)\langle \frac{1}{2}{\bf h}^{-s_k}_{\bf k}, {\bf h}^{s_k}_{\bf k} \rangle dt = \partial_\tau c_{\bf k}^{s_k}\frac{1}{2t}\int_{-\tlim}^\tlim 1 dt = \partial_\tau c_{\bf k}^{s_k}(\tau)$. Note that the term $\partial_\tau c_{\bf k}^{s_k}(\tau)$ can be factored out of the integration over $t$ because $t$ and $\tau$ are independent variables as has been explained above. Also note that the integrand is equal to one because we have used the fact that $\frac{1}{2}\langle {\bf h}^{-s_k}_{\bf k},{\bf h}^{s_k}_{\bf k}\rangle = 1$ as has been explained earlier in section \ref{WAMP}.\\\\
\textbf{{3rd term}}: For the third term we interchange the order of integration, i.e. we swap the operations $\frac{1}{2t}\int_{-\tlim}^\tlim (\cdot )dt $ and $\langle \frac{1}{2} {\bf h}^{s_k}_{\bf k}, \cdot \rangle$. Note that all terms {except} the {exponential term} are functions of $\tau$ (and not $t$) and hence can be factored out of the integration over $t$. So we now concentrate only on the averaging of the exponential term comprising the $\omega$ terms as follows. Recall, for sake of brevity, we use the following notation $\phi(\omega) := (s_k \omega_k - s_p \omega_p - s_q \omega_q)$. Now, in the limit of \textit{large} $t$ or equivalently in the limit of $\omega \tau \ll 1$ (note that $\omega \tau  \sim \frac{\tau}{t}\ll 1 \implies \tau \ll t$ that entails the large limit of $t$ with respect to $\tau$ as explained above), we have
\begin{align}
\frac{1}{2t} \int_{-\tlim}^{\tlim} e^{i\phi(\omega)t} dt  \xrightarrow{t \to \infty}  \krondelta \label{split_int}
\end{align}
where $\krondelta$ is the Kronecker delta function. All other variables besides the exponential term that make up the third term of eq.~(\ref{R-RHD_ham0}) are functions of $\tau$ and upon being operated by $\langle \frac{1}{2} {\bf h}^{s_k}_{\bf k}, \cdot \rangle$, together with the frequency delta term, result in the right hand side of eq.~(\ref{wave_amplitude1}).

{Thus the delta function over $\omega$ {appears naturally} on applying the solvability condition.} This automatically guarantees that the conservation laws of the three wave dispersive system are not violated by the amplitude equation given by eq.~(\ref{wave_amplitude1}). In summary, a necessary separation of scales shown in eq.~(\ref{scale_sep}) has enabled us to average out the exponential term elapsing at slow inertial time scale while retaining the terms that evolve at the advective time scale $\tau$.

\subsection{Energy conservation in wave triads}\label{invariance_proof}
First we show that $\int \frac{1}{k_{\perp}^2}e_{\bf p}e_{\bf q}\delta_{{\bf k},{\bf p}{\bf q}}\delta_{\omega_{k},\omega_{p}\omega_{q}} d{\bf k} = 0$. To show this, we approximate the delta function with a limiting exponential function as: $\delta_{\omega_{k},\omega_{p}\omega_{q}} \approx \lim_{\sigma \to 0}e^{-\frac{\omega_{k}-\omega{p}-\omega_{q}}{\sigma}}$.
\begin{align}
& \int  \frac{1}{k_{\perp}^2}e_{\bf p}e_{\bf q}\delta_{{\bf k},{\bf p}{\bf q}} \delta_{\omega_{k},\omega_{p}\omega_{q}} d{\bf k} \nonumber \\ &\propto \int \delta_{k_z,p_z q_z} \int\frac{1}{k_{\perp}^2}\delta_{k_{\perp},p_{\perp} q_{\perp}}\lim_{\sigma \to 0}e^{-\frac{\omega_{k}-\omega{p}-\omega_{q}}{\sigma}} dk_{\perp}dk_z  \nonumber \\
&= \int \delta_{k_z,p_z q_z} \lim_{\sigma \to 0}\int\frac{1}{k_{\perp}^2}e^{-(\frac{k_z}{k_{\perp}} - \frac{p_z}{p_{\perp}} - \frac{q_z}{q_{\perp}})\sigma^{-1}} \delta_{k_{\perp},p_{\perp} q_{\perp}}dk_{\perp}dk_z \nonumber \\
&= \int \delta_{k_z,p_z q_z} \lim_{\sigma \to 0} e^{(\frac{p_z}{p_{\perp}} + \frac{q_z}{q_{\perp}})\sigma^{-1}}{\int\frac{1}{k_{\perp}^2}e^{-\frac{k_z}{\sigma k_{\perp}}} \delta_{k_{\perp},p_{\perp} q_{\perp}}dk_{\perp}}dk_z \nonumber \\ &= 0.\label{A2_1}
\end{align}
The above equation is zero on account of the integral ${\int\frac{1}{k_{\perp}^2}e^{-\frac{k_z}{\sigma k_{\perp}}} \delta_{k_{\perp},p_{\perp} q_{\perp}}dk_{\perp}}$ being zero as shown below. 
Using integration by parts and the fact that for some arbitrary continuous function $f(x)$, $\int f(x)\delta_{x-a}dx = f(a)$, we have, $I = \int\frac{1}{k_{\perp}^2}e^{-\frac{k_z}{\sigma k_{\perp}}} \delta_{k_{\perp},p_{\perp} q_{\perp}}dk_{\perp}=\frac{1}{k_{\perp}^2}\int e^{-\frac{k_z}{ \sigma k_{\perp}}} \delta_{k_{\perp},p_{\perp} q_{\perp}}dk_{\perp} + \int \frac{2}{k_{\perp}^3} e^{-\frac{k_z \sigma^{-1}}{p_{\perp} + q_{\perp}}} dk_{\perp}=\frac{1}{k_{\perp}^2}e^{-\sigma^{-1}\frac{k_z}{p_{\perp} + q_{\perp}}} - \frac{1}{k_{\perp}^2}e^{-\sigma^{-1} \frac{k_z}{p_{\perp} + q_{\perp}}} = 0$. Next, by following a similar argument it can be shown that $\int \frac{1}{k_{\perp}^2}e_{\bf k}e_{\bf p}\delta_{{\bf k},{\bf p}{\bf q}}\delta_{\omega_{k},\omega_{p}\omega_{q}} d{\bf k} = 0$ and $\int \frac{1}{k_{\perp}^2}e_{\bf k}e_{\bf q}\delta_{{\bf k},{\bf p}{\bf q}}\delta_{\omega_{k},\omega_{p}\omega_{q}} d{\bf k} = 0$.






%
%
%
\section{References}
\bibliographystyle{elsarticle-num-names} 
\bibliography{WT_Sen_PhysD_v1}

\end{document}